%% file: main.tex
\documentclass{jfm}
\usepackage{graphicx}
\usepackage{epstopdf, epsfig}
\usepackage{lineno}
\usepackage{amsmath}

\shorttitle{--}
\shortauthor{Grace, Richter, Berk and Bragg}

\title{Effects of settling on inertial particle slip velocity statistics in wall bounded flows}

\author{Andrew P. Grace\aff{1}
  \corresp{\email{agrace4@nd.edu}},
  David Richter\aff{1}
  \and Tim Berk\aff{2}
 \and Andrew D. Bragg\aff{3}}

\affiliation{\aff{1}Department of Civil and Environmental Engineering and Earth Sciences, University of Notre Dame,
Notre Dame, Indiana 46556, U.S.A.,
\aff{2}Department of Mechanical and Aerospace Engineering, Utah State University, Logan, Utah 84322, U.S.A.,
\aff{3}Department of Civil and Environmental Engineering, Duke University, Durham, North Carolina 27708, U.S.A.}

\begin{document}

\maketitle

\begin{abstract}
Developing reduced order models for the transport of solid particles in turbulence typically requires a statistical description of the particle-turbulence interactions. In this work, we utilize a statistical framework to derive continuum equations for the moments of the slip velocity of inertial settling Lagrangian particles in a turbulent boundary layer. Using coupled Eulerian-Lagrangian direct numerical simulations, we then identify the dominant mechanisms controlling the slip velocity variance, and find that for a range of $\mathrm{St}^+$, $\mathrm{Sv}^+$, and $\mathrm{Re}_\tau$, the slip variance is primarily controlled by local differences between the ``seen" variance and the particle velocity variance, while terms appearing due to the inhomogeneity of the turbulence are sub-leading until $\mathrm{Sv}^+$ becomes large.
We also consider several comparative metrics to assess the relative magnitudes of the fluctuating slip velocity and the mean slip velocity, and we find that the vertical mean slip increases rapidly with $\mathrm{Sv}^+$, rendering the variance relatively small --- an effect found to be most substantial for $\mathrm{Sv}^+>1$. Finally, we compare the results to a model of the acceleration variance \citep{berk_dynamics_2021} based the concept of a response function described in \cite{csanady_turbulent_1963}, highlighting the role of the crossing trajectories mechanism. We find that while there is good agreement for low $\mathrm{Sv}^+$, systematic errors remain, possibly due to implicit non-local effects arising from rapid particle settling and inhomogeneous turbulence. We conclude with a discussion of the implications of this work for modeling the transport of coarse dust grains in the atmospheric surface layer. 
\end{abstract}


\section{Introduction}

The study of the transport of inertial particles through fluids finds numerous applications in the natural sciences and in industry. A significant focus is placed on understanding how small but heavy particles respond to turbulence processes, and how to model these processes in a physically coherent way. One such example is understanding the global transport of coarse dust particles (30-100 $\mu$m) once they are emitted from the surfaces of arid regions \citep{rosenberg_quantifying_2014,meng_improved_2022,adebiyi_review_2023,kok_mineral_2023}. These particles can be lofted high into the turbulent atmosphere where they can be transported many hundreds to thousands of kilometres depending on their size \citep{shao_dust_2008,van_der_does_mysterious_2018}. The interactions between the dust particles and the carrier phase must be parameterized since such interactions occur at the particle scale, and cannot be represented explicitly due to unrealistic computational requirements. Understanding the impacts of turbulence on particle transport characteristics, such as emission and deposition \citep{kok_physics_2012}, will help us to understand their overall role in global climate processes \citep{kok_scaling_2011,ryder_coarse_2019,kok_mineral_2023}, biogeochemical cycles \citep{ryder_coarse-mode_2018}, and human health.

From a dynamical perspective, solid particles are subjected to various forces as they travel through a turbulent flow. For small (relative to the local Kolmogorov scale) and dense (relative to the carrier phase) spherical particles, the most important forces are due to gravity and hydrodynamic drag \citep{maxey_equation_1983}. Since the seminal work of \citet{wang_dispersion_1993}, there has been a significant push to try to understand how gravity and turbulent drag couple together to affect both mean and fluctuating quantities through experiment and simulation \citep{aliseda_effect_2002,good_settling_2014,rosa_settling_2016,tom_multiscale_2019,mora_effect_2021,ferran_experimental_2023}. Importantly, the bias created by gravity leads to a more rapid decorrelation of the turbulence along particle trajectories, meaning that there is an implicit and non-linear coupling between gravity and turbulent drag, resulting in a fundamental change in the forcing induced by turbulence. 
One of the effects of the implicit coupling between gravity and turbulent drag is known as crossing trajectories \citep{yudinemi_physical_1959,csanady_turbulent_1963}, which has been shown to increase the horizontal and vertical components of particle acceleration variance in simulations of settling Lagrangian point particles in HIT \citep{ireland_effect_2016}, in numerical simulations of turbulent boundary layers \citep{lavezzo_role_2010}, as well as laboratory experiments in both setups \citep{gerashchenko_lagrangian_2008,berk_dynamics_2021}.
The turbulent drag is often quantified via the particle slip velocity, which is the difference between the fluid velocity seen by the particle, and the particle's velocity. Understanding the controlling mechanisms of the particle slip velocity and their magnitude within wall bounded turbulence is key for diagnosing a particle Reynolds number \citep{balachandar_scaling_2009}, as well building physically coherent stochastic dispersion models for RANS \citep{arcen_simulation_2009} applications. 
 
In this work, we are particularly focused on understanding the dynamic regimes characteristic of coarse dust particles in Earth's atmospheric surface layer (the lowest 100 metres of Earth's atmosphere). 
Specifically, we consider how gravitational acceleration implicitly modifies the mechanisms controlling the particle slip velocity in a turbulent boundary layer (TBL). In a TBL, turbulence is driven by fluid shear originating at the solid lower boundary, resulting in turbulence inhomogeneity and a height dependence of the parameters governing both the turbulent flow and the particle transport. Dynamically, a turbulent boundary layer is characterized by a very thin laminar sublayer where viscosity plays a dominant role, followed by a smooth transitional layer, known as the buffer layer, to a layer where viscous effects become negligible, known as the logarithmic layer. A review of turbulent boundary layers can be found in \citet{smits_highreynolds_2011}. 

Many studies of particle transport in TBLs tend to ignore the impact of gravitational acceleration \textit{a priori} in an attempt to try to decouple the effects of turbulent drag and gravity \citep{marchioli_statistics_2008,balachandar_scaling_2009,zamansky_acceleration_2011,johnson_turbophoresis_2020}. However, due to the implicit coupling between gravity and turbulent drag, we must take care when extrapolating results from studies without gravity to those with gravity \citep{brandt_particle-laden_2022}. Furthermore, much of our understanding of particle-laden flows under the influence of gravitational settling comes from numerical \citep{good_settling_2014,bec_gravity-driven_2014,ireland_effect_2016,tom_multiscale_2019} or laboratory \citep{aliseda_effect_2002,mora_effect_2021,ferran_experimental_2023} configurations of homogeneous isotropic turbulence (HIT), due to the relative simplicity of the setup. There are a few studies aiming to understand the statistical behaviour of settling inertial particles in turbulent boundary layers \citep{lavezzo_role_2010,lee_effect_2019,berk_transport_2020,bragg_mechanisms_2021,bragg_settling_2021,berk_dynamics_2023}, and while they are far less numerous, they indicate the potential for gravitational settling to modify the dynamics of particle settling and two-way coupling due to the presence of the solid boundary. Indeed, \citet{bragg_settling_2021} showed that gravitational settling can have a strong impact on the particle transport in a TBL even for very small settling numbers for which is has been traditionally assumed that the effect of settling should be negligible. Having quantitative evidence as to when we may apply models designed under the assumptions of homogeneous turbulence to dynamics in a TBL, and when the settling is important for the particle transport, would be a useful starting point when designing a more unified theory. 

In the following work, our goals are: 
\begin{enumerate}
    \item[1.] Derive continuum equations for moments of the particle slip velocity and identify the leading order balance of the variance throughout the turbulent boundary layer. 
    \item[2.] Determine the parametric conditions under which the slip velocity is governed by its mean component, fluctuating component, or some combination of both.
    \item[3.] Compare the results from the DNS in a TBL to a model based on the response function in homogeneous turbulence approach outlined in \cite{csanady_turbulent_1963} and \cite{berk_dynamics_2021}, and identify potential discrepancies.
    \item[4.] Discuss implications for the transport of coarse particles in the atmospheric surface layer.
\end{enumerate}

Section \ref{background} provides the technical background on the carrier and particle phase phase equations as well as the governing parameters. We also derive the diagnostic equation for the vertical component of the slip velocity variance and discuss the model hierarchy. We choose to focus on the slip velocity variance specifically for several reasons. First, accurate estimates of the particle Reynolds number rely on the estimates of the magnitude of the slip velocity, and this quantity is not easily accessible in a laboratory or field setting directly. Second, it is related to the particle acceleration variance, which gives us a clue as to how the particles respond to turbulent structures within the flow. Section \ref{results} presents the results of the study, while section \ref{summary} summarizes and provides a discussion on how the results relate to coarse dust transport in the atmospheric surface layer. 


\section{Technical Background \label{background}}
\subsection{Carrier Phase \label{subsect:carrier_phase}}

\begin{figure}
    \centering
    \includegraphics[width=\textwidth]{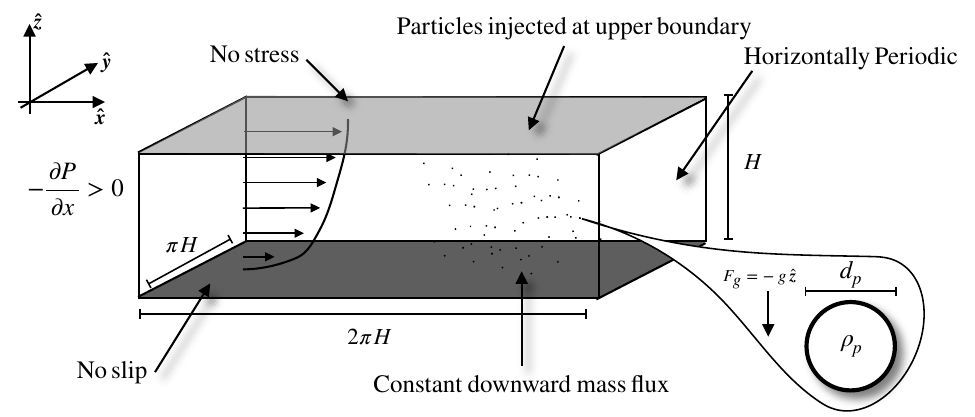}
        \caption{A schematic of the numerical setup. The domain is a rectangular channel of height $H$, streamwise length $2\pi H$ and spanwise width $\pi H$. The flow is periodic in the horizontal and is driven by a constant pressure gradient in the streamwise direction, while the no-stress and no-slip boundary conditions are enforced at the top and bottom boundaries respectively. Particles are injected at the upper boundary at a random horizontal location with an initial velocity equal to the fluid velocity at their location and removed when they contact the bottom boundary. They are allowed to rebound elastically off the upper boundary.}
    \label{fig:schematic}
\end{figure}

In this work, we use the NCAR Turbulence with Lagrangian Particles Model \citep{richter_inertial_2018} to model one-way coupled inertial particles settling through a turbulent boundary layer. This code has been validated and used in multiple studies focused on inertial particle settling and transport in turbulent boundary layers \citep{richter_inertial_2018,wang_inertial_2019,bragg_mechanisms_2021,gao_direct_2023,grace_reinterpretation_2024}. For the carrier phase, we use direct numerical simulations (DNS) to solve the three-dimensional, incompressible Navier-Stokes equations in a turbulent open channel flow setup:
\begin{align}
    \frac{D\boldsymbol{u}}{Dt} &= -\frac{1}{\rho_a}\nabla p + \nu\nabla^2\boldsymbol{u} - \frac{1}{\rho_a}\frac{d P}{dx}\hat{\boldsymbol{x}}, \\
    \nabla \cdot \boldsymbol{u} &= 0.
\end{align}
A schematic of the setup is presented in figure \ref{fig:schematic}.
In the above equations, $\frac{D}{Dt}$ represents the material derivative, $\boldsymbol{u}$ represents the three dimensional flow velocity, $p$ represents the turbulent pressure field, $\rho_a$ is the carrier phase density, and $\nu$ is the kinematic viscosity. As we are primarily focused on wall normal motion in this work, we will refer to fluctuating vertical fluid velocities as $w^\prime$.

At the lower boundary, a no-slip boundary condition is enforced, while at the upper boundary, a no-stress boundary condition ($du/dz = dv/dz = 0$ at $z = H$) is enforced. The domain is periodic in the $x$ and $y$ directions. The background state of the carrier phase is established by accelerating the flow with an imposed pressure gradient, $-dP/dx>0$ (note that $\hat{\boldsymbol{x}}$ is the unit vector in the streamwise direction) and allowing the flow to become turbulent. The magnitude of the pressure gradient allows us to define a friction velocity $u_\tau = \sqrt{\tau_w/\rho_a}$, where $\tau_w$ is the stress at the lower boundary. Using the friction velocity, the height of the domain, $H$, and viscosity of the carrier phase, we can define a friction Reynolds number of $\mathrm{Re}_\tau = \frac{u_\tau H}{\nu}$. Friction Reynolds numbers for each simulation presented in this work can be found in Table \ref{tab:cases}. This setup is identical to that used in \cite{grace_reinterpretation_2024}.

We can define the local Kolomogorov timescale, velocity scale, and acceleration scale, 
\begin{linenomath*}
    \begin{equation}
        \tau_\eta = \left(\frac{\nu}{\epsilon}\right)^{1/2},\quad v_\eta = \left(\nu \epsilon\right)^{1/4},\quad a_\eta = \left(\frac{\epsilon^3}{\nu}\right)^{1/4},
    \end{equation}
\end{linenomath*}
respectively, which represent the smallest relevant length scales of the turbulence. These parameters will be used to characterize the turbulent flow below.
In statistically stationary, homogeneous isotropic turbulence, the above scales are constants, but for wall bounded turbulence they depend on height. Since the turbulence intensity decreases with height outside of the very thin viscous sublayer adjacent to the wall, so too does the kinetic energy dissipation rate resulting in a height variation of the Kolmogorov microscales.  
When the TBL is horizontally homogeneous, the mean dissipation rate $\epsilon$ is a function of the distance from the solid boundary. Within the logarithmic layer, it scales as
\begin{linenomath*}
    \begin{equation}
        \epsilon \sim O\Big(\frac{u_\tau^3}{\kappa z}\Big), \label{dissipation}
    \end{equation}
\end{linenomath*}
where $u_\tau$ is the friction velocity, and $\kappa$ is the von Karman constant. However, when calculating the Kolmogorov timescale, velocity scale, and length scale, we use the dissipation computed in the DNS.

\subsection{Dispersed Phase}
Our main focus is towards on coarse dust transport in the atmospheric surface layer. Dust particles can range in size, but even coarse grains (roughly 30-100 $\mathrm{\mu m}$) are significantly smaller than the local Kolmogorov scale, which can be in the range of several millimetres. Indeed, these particles are also significantly denser than the carrier phase, though their volume fractions can be quite low once they are above the emission layer. With these assumptions in mind, for each particle (the dispersed phase), we apply the point-particle approximation and apply the conservation of momentum for a rigid spherical particle subjected to linear hydrodynamic drag and gravity. Furthermore, as we are concerned with the dilute limit, two-way coupling and particle-particle interactions may be ignored. Since these particles are much denser than the carrier phase, we may also ignore added mass and Basset-History forces. The one-way coupled point particle approach also has the added benefit that each particle is independent from each other particle, effectively removing the volume fraction as a governing parameter. This allows us to increase the number of particles to ensure convergence of the statistics of interest without affecting the flow. 

The equations of motion are 
\begin{eqnarray}
    \frac{d\boldsymbol{v}_p}{dt} =& &\frac{\Psi}{\tau_p}\left(\boldsymbol{u}_f - \boldsymbol{v}_p\right) - \boldsymbol{g}, \label{newton 2} \\ 
    \frac{d\boldsymbol{x}_p}{dt} =& &\boldsymbol{v}_p. \label{part pos}
\end{eqnarray}
Here, $\boldsymbol{v}_p$ is the three dimensional velocity vector for each particle, $\boldsymbol{x}_p$ is the location of each particle in space, $\boldsymbol{g}$ is the gravitational acceleration (which only affects accelerations in the $z$ direction), $\boldsymbol{u}_f$ is the three dimensional instantaneous flow velocity evaluated at the location of the particle. Much of our focus will be placed on the vertical component which we denote as $u_f$ (not bold). $\tau_p$ is the relaxation timescale of the particle, defined as 
\begin{equation}
    \tau_p = \frac{\rho_pd_p^2}{18\rho_a\nu},
\end{equation}
where $\rho_p$ is the particle density and $d_p$ is the particle diameter, and the Stokes settling velocity is defined as $v_g = \tau_p g$. $\Psi = 1 + 0.15\mathrm{Re}_p^{0.687}$ is the Schiller-Neumann correction to the drag force, and $\mathrm{Re}_p$ is the particle Reynolds number. Throughout this work, we change $\tau_p$ by modifying $\rho_p$ (ensuring that $\rho_p\gg\rho_a$) in order to maintain $d_p\ll \eta$. This means that for each case considered, $d_p^+$ (the particle diameter normalized by the viscous scales) is constant across all cases. As we keep the particle diameter fixed for all cases in this work, the particle Reynolds number remains small, meaning that $\Psi\approx 1$. For the theory discussed below in section \ref{phase space}, we follow \cite{bragg_mechanisms_2021} and make the assumption that $\Psi = 1$ for analytical tractability. 

We can now define a set of non-dimensional parameters characterizing the system:
\begin{equation}
    \mathrm{St}^+ = \frac{\tau_pu_\tau^2}{\nu},\quad \mathrm{Sv}^+ = \frac{v_g}{u_\tau},\quad \mathrm{St}_\eta = \frac{\tau_p}{\tau_\eta},\quad \mathrm{Sv}_\eta = \frac{v_g}{v_\eta}. \label{governing params}
\end{equation}
These are the friction Stokes number and the settling velocity parameter based on the viscous scales, and the Stokes number and the settling velocity parameter based on the local Kolmogorov scales. $\mathrm{St}_\eta$ and $\mathrm{Sv}_\eta$ are functions of the distance from the solid boundary, whereas $\mathrm{St}^+$ and $\mathrm{Sv}^+$ are constant parameters. 

Other studies focused on particle settling define other parameters such as a Froude number, $\mathrm{Fr} = a_\eta/g$, \citep{bec_gravity-driven_2014,berk_dynamics_2021}, or a scaled gravity
\begin{equation}
    g^+ = \frac{g\nu}{u_\tau^3},
\end{equation}
which is simply the ratio of $\mathrm{Sv}^+$ to $\mathrm{St}^+$. These parameters are useful as they describe the relative role of gravitational accelerations and turbulent accelerations without referring to $\tau_p$. We will refer to $g^+$ at various points throughout the discussion when necessary. The values for all parameters considered in this work, including the associated ranges of $\mathrm{St}_\eta$ and $\mathrm{Sv}_\eta$, can be found in Table \ref{tab:cases}.


\begin{table}
\centering
\begin{tabular}{ccccccc}
$\mathrm{Re}_\tau$ & $\mathrm{Sv}^+$  & $\mathrm{St}^+$  & $g^+$ & $d_p^+$ & $\mathrm{St_\eta}$ & $\mathrm{Sv}_\eta$ \\ \hline \hline
315 & 0.025   & 10   & $2.5\times 10^{-3}$   & 0.236 & 0.31--4.25  & 0.038--0.14  \\
315 & 0.025   & 50   & $5\times 10^{-4}$     & 0.236 & 1.58--21.25  & 0.038--0.14  \\
 \hline
315 & 0.25    & 10   & $2.5\times 10^{-2}$   & 0.236 & 0.31--4.25  & 0.38--1.4  \\
315 & 0.25    & 50   & $5\times 10^{-3}$     & 0.236 & 1.58--21.25  & 0.38--1.4   \\
\hline 
315 & 0.8    & 10   & $8\times 10^{-2}$     & 0.236 & 0.3--4.29  & 1.22--4.47   \\
\hline 
315 & 2.5     & 10   & $2.5\times 10^{-1}$   & 0.236 & 0.31--4.25  & 3.8--14.0   \\
315 & 2.5     & 50   & $5\times 10^{-2}$     & 0.236 & 1.58--21.25  & 3.8--14.0   \\
\hline
630 & 0.025   & 10   & $2.5\times 10^{-3}$   & 0.236 & 0.24--4.7  & 0.037--0.16  \\
630 & 0.025   & 50   & $5\times 10^{-4}$     & 0.236 & 1.18--23.17  & 0.037--0.16   \\
630 & 0.025   & 100  & $2.5\times 10^{-4}$   & 0.236 & 2.38--46.33  & 0.037--0.16   \\ \hline
630 & 0.25    & 10   & $2.5\times 10^{-2}$   & 0.236 &  0.24--4.7  & 0.37--1.60  \\
630 & 0.25    & 50   & $5\times 10^{-3}$     & 0.236 & 1.18--23.17  & 0.37--1.60   \\
630 & 0.25    & 100  & $2.5\times 10^{-3}$   & 0.236 & 2.38--46.33  & 0.37--1.60   \\ \hline
630 & 0.8     & 10   & $8\times 10^{-2}$   & 0.236 &  0.24--4.7  & 1.17--5.21   \\ \hline
630 & 2.5     & 10   & $2.5\times 10^{-1}$   & 0.236 &  0.24--4.7  & 3.66--16.03   \\
630 & 2.5     & 50   & $5\times 10^{-2}$     & 0.236 & 1.18--23.17  & 3.66--16.03  \\
630 & 2.5     & 100  & $2.5\times 10^{-2}$   & 0.236 & 2.38--46.33  & 3.66--16.03  \\
\hline
1260 & 0.8     & 10   & $8\times 10^{-2}$     & 0.236 & 0.2357--6.29  & 1.0--5.21   \\
\hline

\end{tabular}
\caption{Table of cases discussed throughout the work. Parameter definitions can be found in the main text. The case with $\mathrm{Re}_\tau = 1260$ was run on a $512^3$ grid, all cases with $\mathrm{Re}_\tau  =630$ were run on a $256^3$ grid, while cases with $\mathrm{Re}_\tau  =315$ were run on a $128^3$ grid. \label{tab:cases}}
\end{table}

\subsection{Statistics of inertial particles in a turbulent ASL \label{phase space}}

We adopt the particle phase space approach used in \cite{bragg_mechanisms_2021}, focusing only on the vertical component of the particle equations of motion. First they define the particle PDF in position-velocity space as 
\begin{linenomath*}
    \begin{equation}
        \mathcal{P} = \langle\delta(z_p-z)\delta(w_p-w)\rangle,
    \end{equation}
\end{linenomath*}
which describes the distribution of the vertical components of the particle position and velocity, $z_p(t)$ and $w_p(t)$, in the phase space with coordinates $z$, $w$, and $\langle\cdot\rangle$ represents an ensemble average over all realizations of the system. Note we make frequent use of conditional averages throughout this work, denoted by $\langle\cdot\rangle_{z,w}$ which is short hand for $\langle \cdot | z_p = z,w_p = w\rangle$. We can form an evolution equation for the PDF:
\begin{linenomath*}
    \begin{equation}
        \frac{\partial\mathcal{P}}{\partial t} + \frac{\partial }{\partial z}\left(w\mathcal{P}\right) + \frac{\partial}{\partial w}\left(\langle a_p\rangle_{z,w}\mathcal{P}\right) = 0 \label{continuum}
    \end{equation}
\end{linenomath*}
where we have defined $\langle a_p\rangle_{z,w} = \tau_p^{-1}\left(\langle u_f\rangle_{z,w}-w\right) - g$ as the vertical particle acceleration conditioned on $z_p = z$ and $w_p = w$ based on the vertical component of \eqref{newton 2}.
The utility of this equation comes from the fact that we can derive evolution equations for each moment. Recall that the $n$th moment is defined as
\begin{linenomath*}
    \begin{equation}
        \langle w_p^n\rangle_z = \frac{1}{\varrho}\int_{-\infty}^{\infty}w^n\mathcal{P}dw \label{nth moment}
    \end{equation}
\end{linenomath*}
where the notation $\langle \cdot \rangle_z$ represents an ensemble average conditioned on $z_p = z$. Of importance to the present work will be the first and second moments (i.e. $n=1$ and $n=2$). The evolution equation for the first moment is:
\begin{linenomath*}
\begin{equation}
    \langle w_p\rangle_z = \langle u_f\rangle_z - v_g -\frac{\tau_p}{\varrho}\frac{d}{dz}\varrho\langle w_p^2\rangle_z.  \label{cons momentum}  
\end{equation}
\end{linenomath*}
The details of the derivation of this equation can be found in \citet{bragg_mechanisms_2021}, and the general form of this equation (for the case where $v_g=0$) for arbitrary moments can be found in \cite{johnson_turbophoresis_2020}. Eq. \eqref{cons momentum} says that the average settling velocity comes from the average fluid velocity sampled by the particles, the laminar Stokes settling velocity, and a term that depends on the derivative of the mean square particle velocity which may be important near a solid boundary. For compactness, we can re-arrange \eqref{cons momentum} into a relationship describing how the mean slip velocity varies with height:
\begin{linenomath*}
\begin{equation}
        \langle u_s\rangle_z =  v_g + \frac{\tau_p}{\varrho}\frac{d}{dz}\varrho\langle w_p^2\rangle_z,  \label{mean slip} 
\end{equation}
\end{linenomath*}
where $\langle u_s\rangle_z=\langle u_f\rangle_z -  \langle w_p\rangle_z$.

Our goal is to derive a continuum equation for the slip velocity variance. To do this, we multiply \eqref{continuum} by $w^2$ and integrate over all $w$. The full details of this operation can be found in \cite{johnson_turbophoresis_2020}.
After integrating, we are left with the following relationship:
\begin{linenomath*}
    \begin{equation}
        \frac{d}{dz}\varrho\langle w_p^3\rangle_z - 2\varrho\langle a_pw_p\rangle_z = 0. \label{johnson}
    \end{equation}
\end{linenomath*}
To expand the acceleration-velocity covariance, we expand to get $\langle aw_p\rangle_z = \langle u_fw_p\rangle_z - \langle w_p^2\rangle_z$ and use the fact that  $\langle u_s^2\rangle_z =\langle u_f^2\rangle_z - 2\langle u_fw_p\rangle_z + \langle w_p^2\rangle_z$ to arrive at
\begin{linenomath*}
\begin{equation}
    2\langle aw_p\rangle_z = \frac{1}{\tau_p}\left(\langle u_f^2\rangle_z - \langle u_s^2\rangle_z - \langle w_p^2\rangle_z\right) - 2\langle w_p\rangle_zg \label{accel covariance}
\end{equation}
\end{linenomath*}
Putting \eqref{johnson} and \eqref{accel covariance} together, we get an equation for the mean squared slip velocity:
\begin{linenomath*}
\begin{equation}
    \langle u_s^2\rangle_z = \langle u_f^2\rangle_z - \langle w_p^2\rangle_z -2\langle w_p\rangle_z v_g-\frac{\tau_p}{\varrho}\frac{d}{dz}\varrho\langle w_p^3\rangle_z. \label{cons variance}
\end{equation}
\end{linenomath*}

Eq. \eqref{cons variance} indicates that the mean squared slip velocity variance is a function of the mean squared sampled velocity $\langle u_f^2\rangle_z$, the mean squared particle velocity $\langle w_p^2\rangle_z$, a drift due to the non-zero average vertical velocity $\langle w_p\rangle_z$, and the vertical derivative of the mean cubed particle velocity $\langle w_p^3\rangle_z$, all of which are implicit functions of particle inertia and gravity. 

At this point, an important distinction must be made. Though the above equation is valid for inertial particles settling through a turbulent boundary layer, it should be noted that when particles settle under gravity, the mean squared quantities are not equal to their variances in general. This arises from the non-zero average settling velocity. Thus, to derive a relationship between the variances, we must do a Reynolds decomposition of each term. The details of this process are omitted here, but can be found in Appendix A. The final relationship is 
\begin{linenomath*}
\begin{equation}
    \langle {u_s'}^2\rangle_z = \underbrace{\overbrace{\underbrace{\langle {u_f'}^2\rangle_z - \langle {w_p'}^2\rangle_z}_{(1)} + R_t}^{(2)} + R_g}_{(3)} \label{cons slip var}
\end{equation}
where $R_t$ and $R_g$ are defined as
\end{linenomath*}
\begin{linenomath*}
    \begin{eqnarray}
                R_t =& &-\frac{\tau_p}{\varrho}\frac{d}{dz}\varrho \langle {w_p'}^3\rangle_z \\
                R_g =& &-\frac{\tau_p}{\varrho}\frac{d}{dz}\left(\varrho\langle w_p\rangle^3_z + 3\varrho\langle w_p\rangle_z\langle {w_p'}^2\rangle_z\right)
                + 2\langle w_p\rangle_z\left( \langle u_s\rangle_z -v_g\right), \label{Rg}
    \end{eqnarray}
\end{linenomath*}
respectively.
This model contains many terms and is quite complex, but it can be broken down into three parts, arranged in order of increasing problem complexity, and the level of complexity highlighted by the over and underbraces.

First, grouped under $(1)$, are the terms that would appear for particles settling through homogeneous turbulence. In this limit, the slip velocity variance is determined exclusively by the difference between $\langle {u_f'}^2\rangle_z$ and $\langle {w_p'}^2\rangle_z$. This is the case for particles both with and without gravity, since gravity and inertia implicitly modify these terms. 
Next, grouped under $(2)$, are the terms that appear for particles dispersing vertically through a turbulent boundary layer in the absence of gravity. Note that all terms encompassed by $(1)$ are included in $(2)$, but when considering those terms covered under $(1)$ in the context of a turbulent boundary layer, they gain implicit height dependence since their magnitudes vary with the distance from the boundary. Furthermore, at this level, a new term appears, denoted by $R_t$. This term is proportional to the derivative of the product of the concentration and the particle velocity triple moment and increases with particle inertia. As $\mathrm{Sv^+}\rightarrow 0$, $\langle {w_p'}^2\rangle_z$ approaches $\langle w_p^2\rangle_z$, but $R_t$ remains, regardless.
Finally, by incorporating gravity, the mean particle velocity is no longer zero, leading to a new term grouped under $(3)$, denoted by $R_g$. The quantities composing $R_g$ are explicitly dependent on both the inhomogeneity of the flow through the vertical derivative, and the non-zero particle settling velocity. For clarity of interpretation, the second term on the right hand side of \eqref{Rg} is written in terms of $\langle u_s\rangle_z - v_g$, which we can see from \eqref{mean slip} is identical to $\frac{\tau_p}{\varrho}\frac{d}{dz}\varrho \langle w_p^2\rangle_z$.
In summary, by considering the continuum equations for the first and second moment of the particle velocity, we have been able to derive an equation for the particle slip velocity. We have identified a hierarchy of terms that appear in homogeneous turbulence and TBLs with and without settling (grouped under $(1)$), those that appear in a TBL without settling (grouped under $(2)$), and those that appear in a TBL with settling (grouped under $(3)$). In the following section, we will identify the importance of these terms throughout the turbulent boundary layer.

\section{Results \label{results}}

\subsection{Tendencies governing the slip velocity variance \label{subsect:tendencies}}

\begin{figure}
    \centering
    \includegraphics[width=\textwidth]{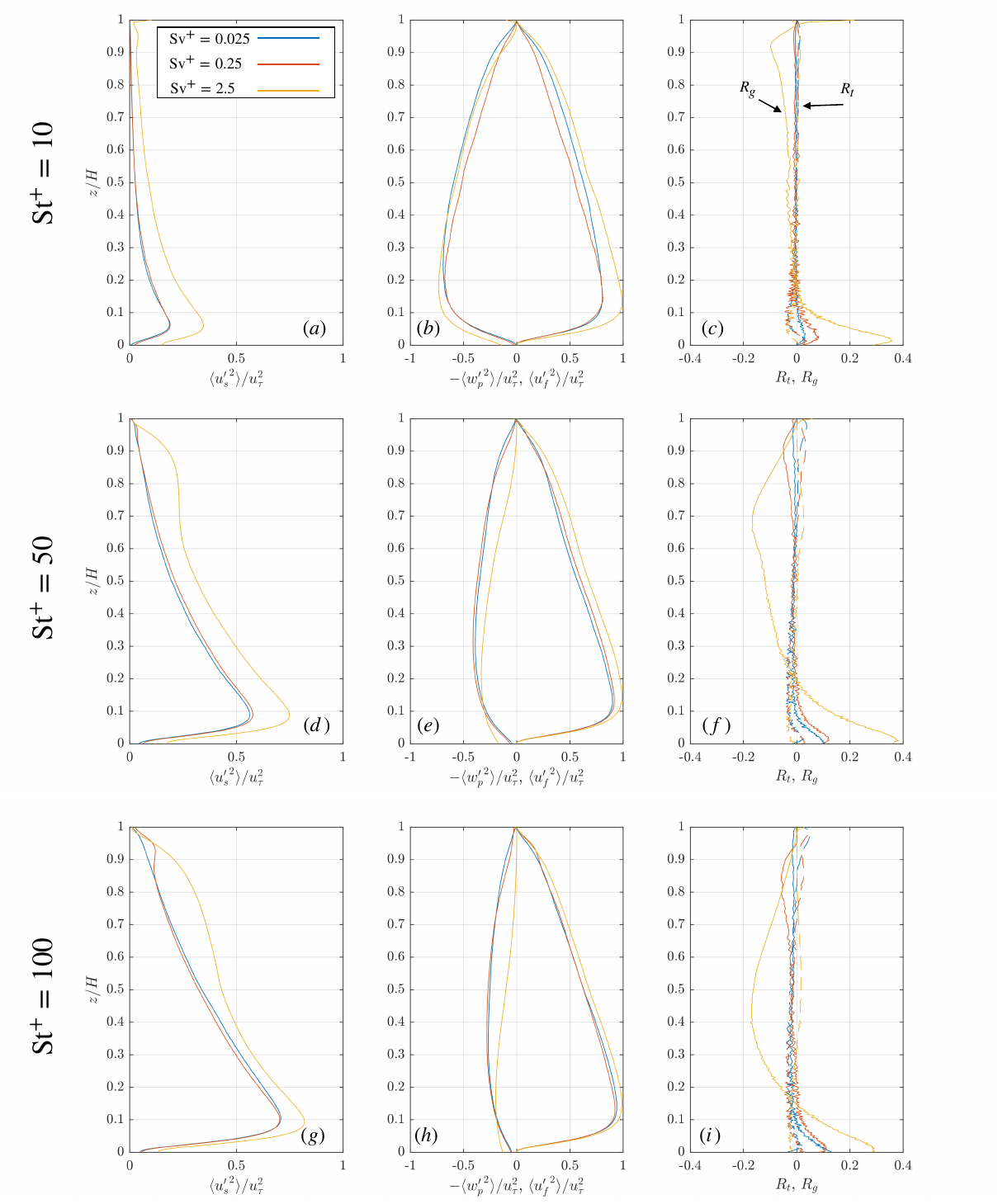}
    \caption{Controlling tendencies for the slip velocity variance according to equation \eqref{cons slip var} at $\mathrm{Re}_\tau=630$. The left column shows the normalized slip velocity variance for each $\mathrm{Sv}^+$, while each row is for a different value of $\mathrm{St}^+$ (shown on the left side of the figure). The centre column shows the (negative) velocity variance and the (positive) seen velocity variance. The right column shows the contributions from $R_t$ and $R_g$. All terms are normalized by $u_\tau^2$.}
    \label{fig:var profs}
\end{figure}
In this section, we consider vertical profiles of $\langle {u_s'}^2\rangle_z$, $\langle {w_p'}^2\rangle_z$, $\langle {u_f'}^2\rangle_z$, $R_t$, and $R_g$. 
Figure \ref{fig:var profs} shows the tendencies in \eqref{cons slip var} for several cases scaled by $u_\tau^2$. Note that all profiles of a given $\mathrm{St}^+$ and $\mathrm{Sv^+}$ in columns two and three of figure \ref{fig:var profs} when added together return the profiles shown in column one, as per \eqref{cons slip var}. An example of this is shown in Appendix C. Each row corresponds to a different friction Stokes number highlighted on the left side of the figure, while each curve in the leftmost and center columns corresponds to a different value of $\mathrm{Sv}^+$. Within the third column, the solid curves correspond to $R_g$ while the dashed curves correspond to $R_t$.
Overall, this figure highlights the dominant tendencies controlling the slip velocity variance as $\mathrm{St}^+$ and $\mathrm{Sv}^+$ are independently changed. 

Generally, the behaviour of the slip variance as a function of the vertical coordinate is qualitatively similar between all cases considered, evident from Figures \ref{fig:var profs}(a), (d), and (g). However, the magnitude of the slip variance for a given case varies throughout the domain, and becomes sensitive to $\mathrm{Sv}^+$ when $\mathrm{Sv}^+$ becomes larger than unity.
Within the logarithmic layer, the slip variance at constant $\mathrm{Sv}^+$ (curves of fixed color) tends to increase rapidly between $\mathrm{St}^+=10$ and $\mathrm{St}^+=50$, but more slowly between $\mathrm{St^+}=50$ and $\mathrm{St}^+=100$. This occurs because the particle velocity variance rapidly decreases in magnitude towards zero, while the variance of the fluid velocity seen by the particle does not, evident by considering figures \ref{fig:var profs}(b), (e), and (h). However, very near the solid boundary (i.e. below $z/H=0.1$), the variance of the fluid velocity seen by the particle approaches zero, while the particle velocity variance remains finite. This is also where $R_t$ and $R_g$  are relatively large, indicating that the primary terms that control the slip variance very near the wall are these terms and negative particle velocity variance.

Likewise, the slip variance at constant $\mathrm{St}^+$ tends to increase most rapidly as $\mathrm{Sv}^+$ surpasses unity. This is due to the fact that particles tend to settle out of locally correlated regions of turbulence faster than they would in the absence of settling, thus experiencing a higher variance in accelerations, and thus their slip velocity. Interestingly, there is some variation in the variance of the fluid velocity seen by the particle with $\mathrm{Sv}^+$. This dependence arises due to the preferential sampling of the fluid velocity field, and the mechanisms responsible for this are essentially the same as those responsible for $\langle u_f\rangle_z$ deviating from zero for an inertial particle (see detailed discussion in \citet{bragg_mechanisms_2021}).


Lastly, the higher order terms ($R_t$ and $R_g$), shown in figures \ref{fig:var profs}(c), (f), and (i), are non-zero but are not leading order within the interior of the domain (note the change in the horizontal scale of these panels), though they are relatively important within the viscous sublayer. Note that above $z/H = 0.1$, these terms are almost completely negligible aside from when $\mathrm{Sv}^+=2.5$. 

In summary, these profiles highlight the fact that within the logarithmic layer, the slip variance is primarily governed by the differences between the variance of the flow velocities sampled by the particles and the particle velocity variance, with contributions coming from $R_g$ when $\mathrm{Sv}^+$ increases beyond unity. However, it is clear that higher order moments of the continuum equations ($R_t$ and $R_g$) may be sub-leading and negligible in most other cases.  It is not until the viscous sublayer where contributions from $R_t$ and $R_g$ become significant in determining the slip variance. Furthermore, the sub-leading behaviour of $R_t$ and $R_g$ in the logarithmic layer does not imply that the inhomogeneity of the turbulence is irrelevant. In fact, the remaining terms ($\langle {u_f'}^2\rangle_z$ and $\langle {w_p'}^2\rangle_z$) may have implicit dependence on the inhomogeneity of the turbulence, and this will be discussed later.

\begin{figure}
    \centering
    \includegraphics[width=\textwidth]{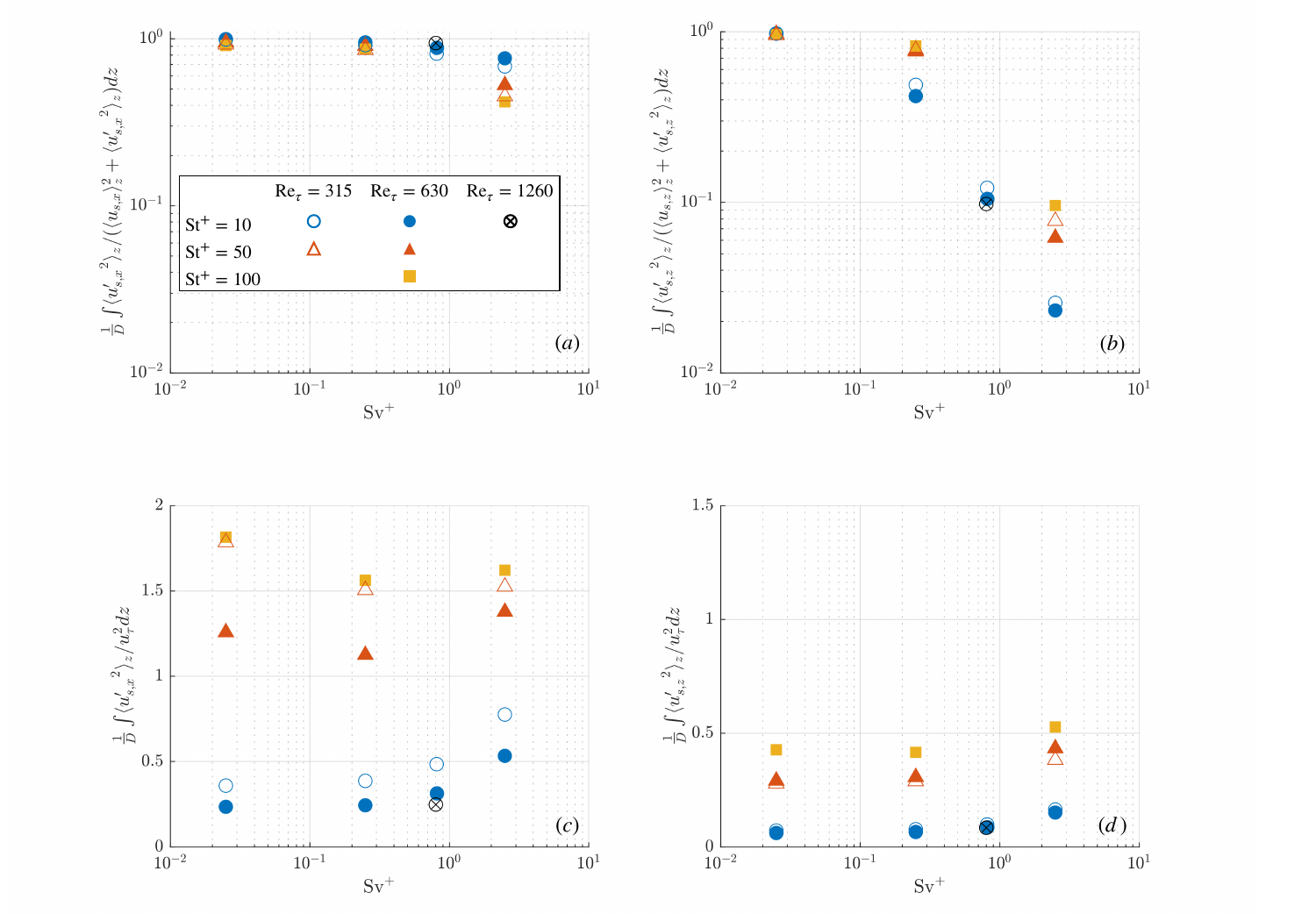}
    \caption{The horizontal and vertical components of $\varphi_r^{(i)}$ (panels (a) and (b)) and $\varphi_s^{(i)}$ (panels (c) and (d)) for all cases in table \ref{tab:cases} plotted against $\mathrm{Sv}^+$. The filled markers correspond to cases at $\mathrm{Re}_\tau=630$, while the open faced markers markers correspond to cases at $\mathrm{Re}_\tau = 315$.}
    \label{fig:fluc ratio}
\end{figure}

As the particles settle, they experience a mean vertical slip velocity according to \eqref{mean slip}, and they also experience a mean horizontal slip due to the background shear.  
Figure \ref{fig:fluc ratio} illustrates the relative contribution of the slip fluctuations to the overall slip velocity by considering two metrics: the integrated relative slip, and the integrated slip variance. These metrics are defined as:
\begin{linenomath*}
    \begin{equation}
        \varphi_r^{(i)} = \frac{1}{D}\int_D\frac{\langle {u_{s,i}^\prime}^2\rangle_z}{\langle {u_{s,i}}\rangle_z^2+\langle {u_{s,i}^\prime}^2\rangle_z} dz,\quad \varphi_s^{(i)} =\frac{1}{D}\int_D\frac{\langle {u_{s,i}^\prime}^2\rangle_z}{u_\tau^2} dz,
    \end{equation}
\end{linenomath*}
respectively, where $D$ is the vertical sub-region of the domain between $z^+=50$ and $z/H = 0.75$, and $i = x,\,z$. These bounds were chosen to eliminate edge effects from the upper boundary condition, though the results are not significantly affected by the choice of the lower bound of the integration. The integrated relative slip helps us to understand the relative importance of the slip fluctuations relative to the mean slip, while the integrated variance provides a simple metric to assess the average slip variance. The integrated relative slip is shown for the horizontal (streamwise) and vertical components of the slip variance (there is no mean slip in the spanwise, so this component is ignored for this discussion) in figures \ref{fig:fluc ratio}(a) and (b), while the components of the integrated slip variance are shown in \ref{fig:fluc ratio}(c) and (d). Each marker style corresponds to a different value for $\mathrm{St}^+$, and the results are plotted against $\mathrm{Sv}^+$ to highlight the role of settling. Filled markers correspond to runs with $\mathrm{Re}_\tau = 630$, empty markers correspond to runs with $\mathrm{Re}_\tau = 315$, and the empty marker filled with an x corresponds to the run with $\mathrm{Re}_\tau = 1260$. Note that since this is an integrated quantity, there is necessarily no information regarding the vertical structure of the profiles. However, an analysis of the profiles themselves (not shown) would provide the same conclusions. 

Figure \ref{fig:fluc ratio} shows that fluctuations of the slip velocity are less important at larger values of $\mathrm{Sv}^+$. For example, for small $\mathrm{Sv}^+$ (independent of $\mathrm{St}^+$), the normalized variance is nearly unity, indicating that the slip variance induced by the turbulent fluctuations are the leading order controller of the overall slip velocity for both the horizontal and vertical components. 
However, we can see that by increasing $\mathrm{Sv}^+$, there is a decrease in the relative slip variance for both components.
The reason for this is clear from an examination of figures \ref{fig:fluc ratio}(c) and (d), which show the integrated slip variance in the horizontal and the vertical respectively. We can see that in a bulk sense, the slip variance in both directions does not strongly vary with $\mathrm{Sv}^+$ (except for perhaps particles with $\mathrm{St^+}=10$ in the horizontal). The implication is that the strong decrease in the relative variance in figure \ref{fig:fluc ratio}(b) (the vertical component) does not come from a decrease in the magnitude of the slip variance itself, but instead a strong increase in the magnitude of the mean slip induced by gravitational settling. Moreover, the same mechanism does not occur in the horizontal slip variance, as the decrease in the normalized slip variance is not nearly as strong.
Furthermore, as $\mathrm{Sv}^+$ increases, particles with $\mathrm{St}^+ =10$ have the largest change in the vertical relative slip variance due to their small slip variance values. This is indicative of the fact that these particles most faithfully follow the flow in the absence of gravity, and as a result, are most sensitive to the growing mean slip as $\mathrm{Sv}^+$ increases.

We also briefly consider a comparison between several Reynolds numbers. There are some slight differences in these metrics as the Reynolds number is varied between 315 and 1260. Here, changes in the relative slip variance are more strongly reflected in the vertical component. As the Reynolds number is increased, the relative slip variance in the vertical, figures \ref{fig:fluc ratio}(b), decreases. Since there are only small variations in the integrated slip variance as the Reynolds number is increased, the change in the integrated relative variance come from changes in the mean slip. The explanation comes from the fact that the the magnitude of $\frac{\tau_p}{\varrho}\frac{d}{dz}\varrho\langle w_p^2\rangle_z$ decreases within the region $D$. It is known from \citet{bragg_mechanisms_2021} that this term is negative within the logarithmic region of the flow, so as it decreases, there is an associated increase in the squared mean slip, $\langle u_s\rangle_z^2$. As the interior of the domain becomes decoupled from the solid boundary as the Reynolds number increases, the relevance of this term becomes diminished, contributing to an overall decrease in the relative slip variance. However, this conclusion is only qualitative as more data at higher Reynolds numbers are necessary to make more quantitative conclusions.

The conclusion of figure \ref{fig:fluc ratio} is that while the overall slip magnitude in the horizontal may be controlled by the slip fluctuations, the mean may end up being the main controller of the slip in the vertical at large $\mathrm{Sv}^+$ and small to moderate $\mathrm{St}^+$, but the relative importance of the mean may be impacted by the Reynolds number of the flow.

\subsection{Relationship to the acceleration statistics}


The slip velocity statistics are directly related to the acceleration statistics of the particles. By considering the acceleration statistics, we can gain an understanding of how strongly gravity implicitly modifies the drag felt by the particles as they traverse the TBL. Moreover, the acceleration variance often gives a clue regarding the turbulent structures which particles are interacting with. For example, \cite{yeo_near-wall_2010} showed that the elongated tails of fluid particle accelerations within the buffer layer and viscous sublayer are due to the vortical structures impinging on the viscous sublayer. \cite{lavezzo_role_2010} attributed particles settling through these same vortical structures to the increase in spanwise and streamwise acceleration variance. In a two-way coupled turbulent Couette flow, \citet{richter_momentum_2013} found that particles tend to damp vertical fluctuations of the near wall dynamics, suggesting a complex feedback cycle.

In the limit of $\mathrm{g}^+\rightarrow 0$ (recall that $g^+=\mathrm{Sv}^+/\mathrm{St}^+)$, or when gravitational accelerations are ignored \textit{a priori}, \citet{bec_acceleration_2006} demonstrated that particles tend to cluster in strain dominated regions of the flow for low Stokes number (i.e. $\mathrm{St_\eta}\lesssim 0.3$), leading to a decrease in the acceleration variance of the particles. At larger Stokes number, the acceleration variance continues to decrease, but is instead due to inertial filtering; particles can no longer respond to turbulent fluctuations with timescales greater than $\tau_p^{-1}$. Ultimately, both processes work to reduce the acceleration variance, but for different reasons \citep{bragg_relationship_2015}.
However, by introducing gravity, particles can settle out of strain dominated regions of the flow, which may actually contribute to an increase in their acceleration variance, and this often referred to as the crossing trajectory mechanism \citep{csanady_turbulent_1963}.
\citet{berk_dynamics_2021} and \citet{ireland_effect_2016-2} showed that the importance of the crossing trajectories mechanism on the acceleration statistics is due to both $\mathrm{St}_\eta$ and $\mathrm{Sv}_\eta$ (or alternatively $1/\mathrm{Fr} = g/a_\eta$). They showed that for large $g/a_\eta$ (equivalent to large $g^+$ in  our context), gravitational accelerations become increasingly important to the dynamics, leading to a peak in the acceleration variance at sufficiently high $g/a_\eta$, around $\mathrm{St}_\eta =\mathcal{O}(1)$. 
In the following results, we highlight some similarities of the computed slip and acceleration variance to results from \citet{berk_dynamics_2021}, who focused on modeling the slip and acceleration variance in homogeneous turbulence in terms of the variance of the fluid along the particle trajectory. Our goal is to compare our model results in a turbulent boundary layer to the predictions of their model for homogeneous turbulence (derived in Appendix B). 

\begin{figure}
     \centering
     \includegraphics[width=\textwidth]{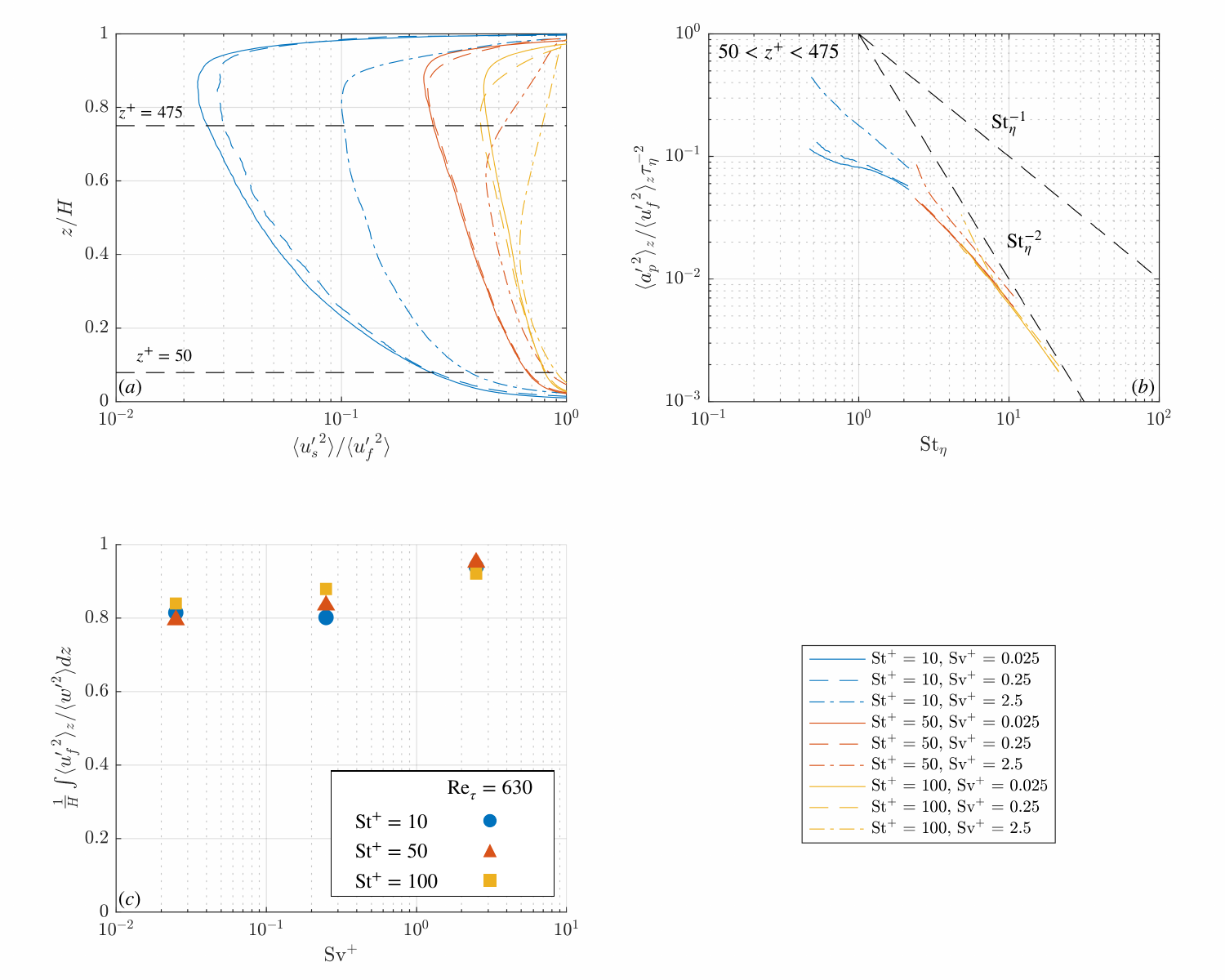}
     \caption{Panel (a) shows the slip variance normalized by the seen variance for all cases in table \ref{tab:cases} at $\mathrm{Re}_\tau = 630$. The horizontal dashed lines denote heights of $z^+=50$ and $z/H = 0.75$. Panel (b) shows the normalized acceleration variance over the same range plotted against the local value of $\mathrm{St}_\eta$, while the dashed line represents the $\mathrm{St}_\eta^{-2}$ scaling. The colors of each curve in panels (a) and (b) correspond to values of $\mathrm{St}^+$, while the line styles correspond to values of $\mathrm{Sv}^+$. Panel (c) shows ratio of the seen variance to the unconditional variance averaged over the entire vertical extent plotted against $\mathrm{Sv}^+$ for all cases at $\mathrm{Re}_\tau = 630$.}
     \label{fig:accel_stats}
\end{figure}

First, we consider profiles of the relative slip variance, which we define as $\langle {u_s^\prime}^2\rangle_z/\langle {u_f^\prime}^2\rangle_z$, shown in figure \ref{fig:accel_stats}(a). Here we again focus on the region $D$, which is the region between $z^+=50$ and $z/H = 0.75$ in order to omit effects from viscous sublayer and the upper boundary condition respectively (denoted by black dashed lines on the figure). The general trend is that by increasing $\mathrm{Sv}^+$ at a given $\mathrm{St}^+$, the relative slip variance tends to increase, with the most dramatic increase coming as $\mathrm{Sv}^+$ is increased beyond unity, which is probably a reflection of the crossing trajectories mechanism. Furthermore, we can see that relative change between $\mathrm{Sv}^+=0.25$ and $\mathrm{Sv^+}=2.5$ decreases as $\mathrm{St}^+$ increases. The reason for this is discussed more below.

We can also consider the relative acceleration variance, $\langle {a_p^\prime}^2\rangle_z/\langle {u_f^\prime}^2\rangle_z\tau_\eta^{-2}$, plotted against the local value of $\mathrm{St}_\eta$, shown in figure \ref{fig:accel_stats}(b). We can see that as the range of $\mathrm{St}_\eta$ increases (by changing $\mathrm{St}^+$), the relative acceleration variance approaches the asymptotic relationship $\mathrm{St}_\eta^{-2}$. However, for moderately inertial particles, characterized by the $\mathrm{St}^+=10$ cases, there is more potential for gravity to increase the relative acceleration variance. For example, due to the crossing trajectories mechanism, we can see that when $\mathrm{Sv}^+ = 2.5$ for these particles, the acceleration variance is much larger, and tends to scale as $\mathrm{St}_\eta^{-1}$, which is consistent with the results from \cite{balachandar_scaling_2009} when $\tau_\eta \ll \tau_p \ll \tau_{l,p}$ and from \cite{berk_dynamics_2021} in roughly the same range of $\mathrm{St}_\eta$. 
From figure \ref{fig:accel_stats}(b), we can see that the crossing trajectories mechanism is not strong for large $\mathrm{St^+}$ particles, since these particles approach the asymptotic $\mathrm{St}_\eta^{-2}$ scaling across the entire TBL. The implication here is that extremely inertial particles tend not to respond to high frequency and intermittent turbulent fluctuations associated with changes in the sampled fluid environment anywhere across the TBL.
However, when particles are moderately sized, such that they achieve $\mathrm{St}_\eta \sim 1$ there is a region within the logarithmic layer of the TBL where they become susceptible to crossing trajectory effects, and this leads to an increase in their relative acceleration variance.

As a final point on this discussion, a potential shortcoming of analyzing the relative slip variance and the relative acceleration variance is that they are written in terms of the fluid velocity variance along the particle trajectory, which is an unknown quantity \textit{a priori}. We can relate this to the unconditional variance, for which there are well known models (see \citet{kunkel_study_2006} for example). In figure \ref{fig:accel_stats}(c) we show the vertically integrated ratio of the vertical components of the fluid velocity variance along particle trajectories to the unconditional fluid velocity variance against $\mathrm{Sv}^+$. This integrated ratio approaches unity as $\mathrm{Sv}^+$ increases implying that the seen variance approaches the unconditional variance in this limit. Moreover, this ratio is no less than roughly 0.8 for our range of parameters, implying an acceptable correspondence between these two quantities. This is significant for modeling purposes as our results show that the fluid velocity variance along the particle trajectory may be substituted for the seen variance in a turbulent boundary layer without incurring significant error, having implications for the predictive power the particle statistics in a TBL. \citet{berk_dynamics_2021} also arrived at this conclusion in homogeneous turbulence.

\begin{figure}
    \centering
    \includegraphics[width=\textwidth]{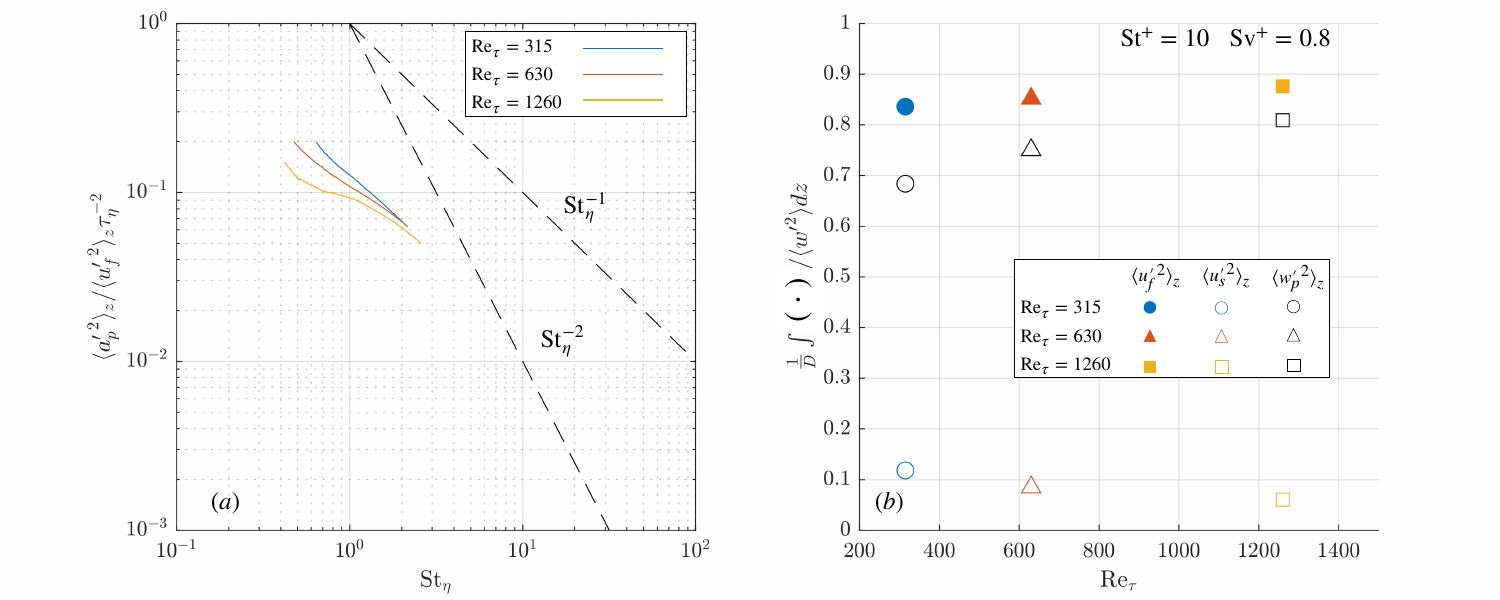}
    \caption{Panel (a) shows the normalized acceleration variance for cases with $\mathrm{St}^+=10$ and $\mathrm{Sv}^+=0.8$ at three different Reynolds numbers as a function of $\mathrm{St}_\eta$. Panel(b) shows the seen variance (filled markers), slip variance (colored empty markers), and particle velocity variance (black markers) normalized by the unconditional variance integrated over the range $D$ as a function of $\mathrm{Re}_\tau$.}
    \label{fig:Re_com}
\end{figure}

We can also consider the impact of Reynolds number on the relative acceleration variance, as well as the averaged components of \eqref{cons slip var}. We can see in figure \ref{fig:Re_com}(a) the relative variance decreases as a function of $\mathrm{St}_\eta$, as we would expect based on figure \ref{fig:accel_stats}, but it is interesting that increasing $\mathrm{Re}_\tau$ further decreases the relative acceleration variance. The reason for this is shown in figure \ref{fig:Re_com}(b). Both the the averaged fluid velocity variance seen by the particles (filled markers), and the particle velocity variance (empty markers; black outline) increase with $\mathrm{Re}_\tau$, but since the particle velocity variance increases faster with $\mathrm{Re}_\tau$, the net effect is a decrease in the slip variance with increasing Reynolds number (and consequently the relative acceleration variance since they are related through $\tau_p$). The increases in $\langle {u_f^\prime}^2\rangle_z$ and $\langle {w_p^\prime}^2\rangle_z$ with Reynolds number when averaged across the domain is probably due to the increasing size of the quasi-homogeneous region of the flow \citep{kunkel_study_2006}. Models of the unconditional vertical fluid velocity variance suggest that this quantity asymptotically approaches a constant in the limit of high Reynolds number. Though our Reynolds numbers are still quite low with regards to those found in the atmospheric surface layer, this notion can still provide some guidance to interpreting our data. As both $\langle {u_f^\prime}^2\rangle_z$ and $\langle {w_p^\prime}^2\rangle_z$ are related to the unconditional fluid velocity variance in some way, we expect that they should follow this behaviour, at least qualitatively. Finally, $R_g$ and $R_t$ are non-zero within the interior of the domain, but their magnitude, discussed later, is secondary to both $\langle {u_f^\prime}^2\rangle_z$ and $\langle {w_p^\prime}^2\rangle_z$ when taking the average across the logarithmic layer (recall the $D$ does not include the viscous sublayer). 

\begin{figure}
     \centering
     \includegraphics[width=\textwidth]{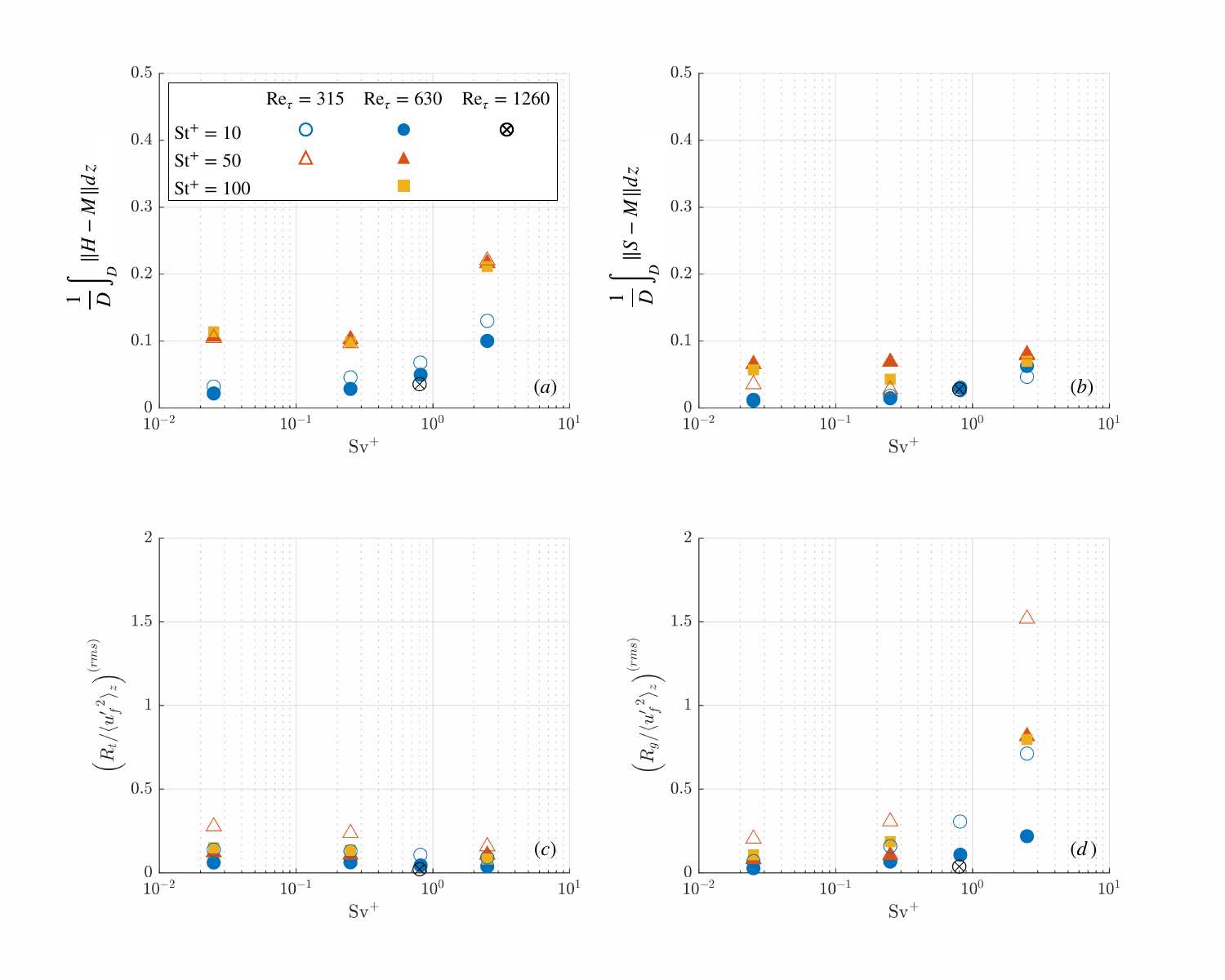}
     \caption{Shown in panels (a)--(d) are $D_1$, $D_2$ and the RMS values of $R_t$ and $R_g$ normalized by the seen variance for all cases in Table \ref{tab:cases}, respectively. Open faced markers represent cases with $\mathrm{Re}_\tau = 315$, while filled markers represent cases with $\mathrm{Re}_\tau = 630$.}
     \label{fig:rms_RtRg}
\end{figure}

To conclude this section, we comment on the applicability of the model proposed by \citet{berk_dynamics_2021} for $\langle {u_s^\prime}^2\rangle_z$ (which we will refer to as BC2021 within the text). A sketch of the derivation of their model is presented in Appendix B. In short, they invoke an argument from \citet{csanady_turbulent_1963} to relate the particle velocity variance to the seen fluid energy spectrum. They then use the fact that the fluid energy spectrum along the particle trajectory is the Fourier transform of the auto-correlation of the fluid velocity along the particle trajectory, which they represent using a two timescale model derived by \citet{sawford_reynolds_1991}. The auto-correlation function involves the decorrelation timescale of the turbulence along the particle trajectory, $\tau_{l,p}$, for which they use the model derived in \cite{csanady_turbulent_1963}. This model assumes that the turbulence is homogeneous and isotropic, and thus particles experience no spatial change in the statistics of the turbulence along their trajectory. Our goal is to compare how the computed relative slip variance within the TBL compares with the modeled slip variance in BC2021. As BC2021 was developed under the assumption of homogeneous turbulence, we can extend it to a TBL by making a locally homogeneous approximation, meaning that any change the slip variance with height is occurs due to local changes in the turbulent dissipation. We first consider \eqref{cons slip var} normalized by the seen variance:
\begin{linenomath*}
    \begin{equation}
        \frac{\langle {u_s^\prime}^2\rangle_z}{\langle {u_f^\prime}^2\rangle_z} = \overbrace{\underbrace{1 - \frac{\langle {w_p^\prime}^2\rangle_z}{\langle {u_f^\prime}^2\rangle_z}}_{\text{$H$}} + \frac{R_t}{\langle {u_f^\prime}^2\rangle_z} + \frac{R_g}{\langle {u_f^\prime}^2\rangle_z}}^{\text{$S$}}. \label{rel slip var}
    \end{equation}
\end{linenomath*}
As we have discussed previously, our analysis will be limited to the logarithmic region of the flow. We also comment (but do not show) that the variance predicted by B2021 is smaller in magnitude than the slip variance computed by the DNS throughout the this region. 

As the model developed in \citet{berk_dynamics_2021} was for settling particles in homogeneous turbulence, the slip variance in that case was only due to the difference between $\langle {u_f^\prime}^2\rangle_z$ and $\langle {w_p^\prime}^2\rangle_z$ (see Appendix B). We denote this subset of terms in \eqref{rel slip var} by $H$ after normalizing by $\langle {u_f^\prime}^2\rangle_z$. Therefore, our first comparison is between $H$ and the BC2021 model (which we denote by $M$ for brevity); mathematically, we consider
\begin{linenomath*}
    \begin{equation}
        D_1 = \frac{1}{D}\int_D\|H - M\| dz,
    \end{equation}
\end{linenomath*}
which is shown in figure \ref{fig:rms_RtRg}(a).

This metric highlights a disparity between the DNS and BC2021, but the differences amount to no more than roughly $0.3\langle {u_f^\prime}^2\rangle_z$ (note both $H$ and $M$ have a factor of $\langle {u_f^\prime}^2\rangle_z$ in their denominator). This discrepancy may come from several sources; one may be due to the fact that the underlying statistics of the turbulent flow change along the particle's trajectory due to the presence of the wall. This behaviour is reflected in the general increase of $D_1$ as $\mathrm{Sv}^+$ increases, and will be discussed more in section \ref{summary}. It is also interesting to note that there are differences associated with $\mathrm{St}^+$, and these differences tend to plateau at large $\mathrm{St}^+$. 


Moreover, there are likely to be differences between the DNS data and BC2021 associated with the relatively small Reynolds numbers considered in this study. Within BC2021, there are several sub-models required to calculate characteristic parameters (i.e. $C_0$ and $a_0$; see Appendix B) of the turbulence, and there may be an associated error incurred when the Reynolds number is low (for example, see the discussion in \citet{lien_kolmogorov_2002}). However, we can see that by increasing $\mathrm{Re}_\tau$, the differences between the DNS and BC2021 tend to decrease suggesting a correspondence at large enough Reynolds number.

Now, we consider how BC2021 compares to the full right hand side of \ref{rel slip var}, which includes $R_t$ and $R_g$ (note that $R_t$ and $R_g$ cannot appear in BC2021 \textit{a priori} due to their assumption of homogeneous turbulence). We will denote this as $S$ and consider the metric 
\begin{linenomath*}
    \begin{equation}
        D_2 = \frac{1}{D}\int_D\|S - M\| dz,
    \end{equation}
\end{linenomath*}
which is plotted in figure \ref{fig:rms_RtRg}(b).

It is apparent that the inclusion of $R_t$ and $R_g$ actually works to decrease differences between the DNS data and BC2021. However, as we established \ref{fig:var profs}(c), (f), and (i), $R_t$ is negligible within the logarithmic region, and $R_g$ is primarily negative, causing a reduction of the relative slip variance. As we noted previously, $M$ underestimates the computed relative slip variance within the logarithmic region of the turbulence, and the result is a decrease in the differences between the BC2021 and the DNS data. Moreover, by considering the root-mean-squared of the normalized $R_t$ and $R_g$, shown in figures \ref{fig:rms_RtRg}(c) and (d), we can see that $R_g$ is much more important than $R_t$. Again, the importance of this term appears when both $\mathrm{St}^+$ and $\mathrm{Sv}^+$ are large, but decreases significantly as $\mathrm{Re}_\tau$ increases.

In summary, BC2021 gives a reasonable estimate for the DNS data in the logarithmic region of the flow. However, at large $\mathrm{St^+}$ and $\mathrm{Sv}^+$, the size of the differences increases, and this is due to combination of the changing statistics of the turbulence along the particle trajectories, as well as the low Reynolds numbers in the DNS, and how these are represented within BC2021. 
However, an important conclusion is that at large Reynolds number, we expect correspondence between the DNS data and BC2021, evidenced by the fact that the differences between BC2021 and the DNS decrease in this limit. Moreover, outside of the viscous sublayer, the importance of $R_t$ and $R_g$ will also be reduced as $\mathrm{Re}_\tau$ increases, meaning correspondence between predictions by the BC2021 model and the variance measured in a TBL will become stronger in this limit. These results suggest that when $\mathrm{St}^+$ and $\mathrm{Sv}^+$ are not too large, we can estimate the slip variance in the logarithmic region of the flow by using BC2021 without incurring significant error.

\section{Summary and Discussion \label{summary}}
\subsection{Summary}

Motivated by coarse particle transport in the atmospheric surface layer, we used coupled Eulerian-Lagrangian simulations to simulate the dynamics of ensembles of inertial particles in boundary layer turbulence. We examined the impact of particle inertia and settling on the mean and fluctuating particle slip velocity. We adapted a mathematical model discussed in \citet{bragg_mechanisms_2021} and \citet{johnson_turbophoresis_2020} for the slip velocity variance for settling inertial particles in a turbulent boundary layer and highlighted the controlling factors throughout the domain. 
We showed that to leading order, the slip variance above of the viscous sublayer was determined by the difference between the seen variance, $\langle {u_f^\prime}^2\rangle_z$, and the particle velocity variance $\langle {w_p^\prime}^2\rangle_z$, except for the largest value of $\mathrm{Sv}^+$, where all terms in \eqref{cons slip var} became comparable. Consequently, as changes in the seen variance were relatively small, changes in the slip variance within the logarithmic layer were primarily governed by a decrease of the particle velocity variance, which was implicitly a function of particle inertia and the particle settling velocity (more on this below). Within the viscous sublayer, the balance became more complicated.  In all cases, $\langle {u_f^\prime}^2\rangle_z$ tended towards zero to adhere to the no-slip condition enforced at the bottom boundary. However, the slip variance remained finite as the particles tended towards $z^+ = 0$ as $\langle {w_p^\prime}^2\rangle_z$, $R_t$, and $R_g$ remained finite. The higher order terms tended to peak within this layer, and the magnitude of the peak tended to increase with $\mathrm{Sv}^+$.
However, by using domain averages, we demonstrated that the relative magnitude of the higher moment terms tended to decrease as the Reynolds number increased, reflecting the fact that at higher Reynolds number, the viscous sublayer becomes much thinner, leading to a smaller contribution when averaged across the domain. As discussed above, these terms may still be important within the viscous sublayer depending on particle parameters, though. 

We also showed that the fluid velocity variance along the particle trajectories exhibited only small changes with $\mathrm{St}^+$, $\mathrm{Sv}^+$ and $\mathrm{Re}_\tau$, and was approximately 80\% of the unconditional fluid velocity variance when averaged across the domain. The differences between the seen and unconditional variances are largest at the smallest $\mathrm{Sv}^+$ considered in this work, though the differences are still relatively small. These differences are likely a result of the relatively large spectrum of turbulent motions at high Reynolds number, and the impact of crossing trajectories, which works to implicitly affect the seen variance (see the discussion in \citet{csanady_turbulent_1963}, for example). This conclusion is quantitatively consistent to the conclusions presented in \citet{berk_dynamics_2021} for their laboratory experiments in homogeneous turbulence.
This correspondence is significant as the seen variance is not known \textit{a priori}. There exist approaches to modeling this quantity \citep{Pozorski_lagrangian_1998,minier_pdf_2001}, but the results of our work suggest that even in a turbulent boundary layer where there is spatial dependence in the of the turbulent quantities, we can approximate the fluid variance along the particle trajectories with the unconditional fluid variance without introducing significant errors. While this reduces the overall accuracy of the slip variance estimate, it increases the predictive power at the field scale, as models for the unconditional variance, such as those discussed in \citet{kunkel_study_2006}, can be employed. 

To examine the relative importance of the fluctuating and mean slip, we considered the ratio of the slip variance to the total mean squared slip velocity, $\langle {u_s}^2\rangle_z$ (i.e. the square of the mean plus the fluctuation). We found that relative to the mean, the vertical slip variance decreased much faster than the horizontal as $\mathrm{Sv}^+$ was varied. However, for both components, we showed that the overall magnitudes in the average sense did not change significantly, indicating that at relatively large $\mathrm{Sv}^+,$ the mean slip was the determining factor in the vertical, while the fluctuating slip was the determining factor in the horizontal. 
This effect was also accentuated for the smallest particles considered, due to their relatively small slip variance. To further complicate the behaviour, the relative size of the slip variance tended to decrease as the Reynolds number was increased, though due to computational restrictions, we can only provide limited guidance on this issue.

We also compared the slip variance computed by the DNS to a model derived for homogeneous isotropic turbulence by \citet{berk_dynamics_2021} (as no such model currently exists for a turbulent boundary layer and is the focus of future work). The main conclusion is that the globally averaged differences (in the absolute sense) were relatively small, but the higher moment terms act as a confounding factor to reduce differences between the model and the DNS data. Thus, care should be taken when extrapolating results from low Reynolds number DNS in turbulent boundary layers to higher Reynolds number experiments in homogeneous turbulence. However, as we know the size of higher moment terms tends to decrease as Reynolds number increase when integrated across the domain, we expect that DNS at higher Reynolds number should tend towards the results derived in \citet{berk_dynamics_2021} outside of the thin viscous sublayer. 

Additionally, due to the inhomogeneous nature of the turbulence, non-local effects implicit to $\langle {u_f^\prime}^2\rangle_z$ and $\langle {w_p^\prime}^2\rangle_z$  may occur at large $\mathrm{Sv}^+$ and $\mathrm{St}^+$. Isolating the importance of non-local effects in a TBL is the focus on ongoing research (for a recent example for a model of $\langle {w_p^\prime}^2\rangle_z$ in a TBL, see \citet{zhang_asymptotic_2023} and references therein), but incorporating them in a model is beyond the scope of the current article. These effects may arise due to the fact that the statistics of the turbulence may change significantly as the particle travels vertically. For example, consider the distance a settling particle travels over one relaxation time: $\delta \sim  |\tau_p\langle w_p\rangle_z|$, where $\langle w_p\rangle_z$ is the average particle settling velocity conditioned on a height $z$ given by \eqref{cons momentum}. In order for the particle trajectory to be altered, turbulent fluctuations must be correlated over this distance. However, if this distance is comparable to the distance over which the characteristics of the turbulence change, then we expect that the particle feels the inhomogeneous nature of the flow. To formalize this quantitatively, consider the local turbulent kinetic energy at a height $z$, $k$. Taylor expanding about this point and truncating after the second term, we have
\begin{linenomath*}
\begin{equation}
    k(z-\delta) = k(z)- \left.\delta\frac{dk}{dz}\right|_z. \label{taylor}
\end{equation}
\end{linenomath*}
To make a locally homogeneous approximation about the turbulence, we must have that 
\begin{linenomath*}
    \begin{equation}
       k(z)\gg \left.\delta\frac{dk}{dz}\right|_z,
    \end{equation}
\end{linenomath*}
i.e. the kinetic energy in a small neighborhood about $z$ (defined by the distance $\delta$) is primarily defined by the kinetic energy measured at a height $z$.
Using \eqref{taylor}, we can estimate under what conditions a locally homogeneous approximation would be appropriate by looking for cases where the second term is small compared to the first. By assuming that the gradient of the turbulent kinetic energy scales as $u_\tau^2z^{-1}$ (also implying $k$ can be scaled by $u_\tau^2$) \citep{smits_highreynolds_2011} we have that $z\gg \delta$.
By normalizing both sides of this inequality by the root-mean-squared turbulent velocity, $u^\prime = k^{1/2}$, we can write $\delta$ in terms of the sum of the settling enhancement, $E = \frac{\langle w_p\rangle_z + v_g}{u^\prime}$ and $v_g$,  \citep{good_settling_2014,loth_particles_2023} as
 \begin{linenomath*}
    \begin{equation}
        \tau_p\left|E + \frac{v_g}{w^\prime}\right| \ll \frac{z}{u^\prime}.
    \end{equation}
\end{linenomath*}
Now normalizing by $\tau_{\eta}$, and observing that $u^\prime  \sim u_\tau$, we can simplify both side of this inequality to reveal that 
\begin{linenomath*}
    \begin{eqnarray}
        \mathrm{St}_\eta\left|E + \mathrm{Sv}_\ell\right| \ll \left(\frac{zu_\tau}{\nu}\right)^{1/2}, \label{inequality}
    \end{eqnarray}
\end{linenomath*}
where we have used the dissipation scaling in \ref{dissipation} to relate $\tau_\eta$ to the vertical coordinate, $z$.
This relationship indicates that both particle inertia and gravity have an explicit role, and an implicit role (through $E$) to play in potential non-local effects.

We know from \citet{good_settling_2014} and \citet{loth_particles_2023} that in small scale laboratory experiments, $E\sim 0.2$ as $\mathrm{St}_\eta$ and $\mathrm{Sv}_\ell$ approach 1, but as both of these parameters increase, $E$ tends back towards zero, and may even become negative \citep{ferran_experimental_2023}. Therefore, for the coarse particles we are concerned with in this work, we can make a locally homogeneous approximation when $\mathrm{St_\eta Sv_\ell}\ll\left(\frac{zu_\tau}{\nu}\right)^{1/2}$. This may not particularly restrictive for the atmospheric surface layer as the Reynolds numbers are $\mathcal{O}(10^6)$, but for laboratory experiments, the integral scales tend to scale with the size of the experimental domain (i.e. $z\sim h$ where $h$ could be the half-height of a channel). This could present a problem making a locally homogeneous approximation for coarse particles in a wind tunnel setup.

For the DNS presented in this work, the above relationship shows that non-local effects are likely only important for cases when both $\mathrm{St^+}$ and $\mathrm{Sv}^+$ are large, as $\mathrm{Sv}_\ell$ tends to scale with $\mathrm{Sv}^+$ since $u^\prime\sim u_\tau$. For example, if we consider cases with $\mathrm{St^+}=100$ and $\mathrm{Sv}^+ = 2.5$ (and assume $E\approx 0$), we can see immediately that \ref{inequality} is not satisfied. This may explain the differences between the DNS and the model in \citet{berk_dynamics_2021} in figure \ref{fig:rms_RtRg}(a) and (b) for these particles. 
One of the main conclusions of this work is that for the governing continuum equation for the particle slip velocity in a turbulent boundary layer, there are tendencies that arise due to the inhomogeneities in the turbulence associated with the presence of the wall and the fact that the particle settling velocity is non-zero. However, as we have shown, for moderately sized particles (characterized by $\mathrm{St}^+$ or $\mathrm{St}_\eta$), and $\mathrm{Sv}^+<1$, these terms are subleading outside of the viscous sublayer. Moreover, the magnitudes of these terms in the logarithmic layer tend to diminish as $\mathrm{Re}_\tau$ increases. Thus, outside of the viscous sublayer, and at moderate local $\mathrm{St}_\eta$ and $\mathrm{Sv}^+$, we can extend models designed for homogeneous turbulence (like that described in \citet{berk_dynamics_2021}) to a TBL, where we must interpret the model as local to a height $z$. However, outside of this regime (i.e. for very large and strongly settling particles), there are implicit non-local effects that appear as particles tend to settle through the flow due to the vertical variation of the turbulent statistics along the particle trajectories.

\subsection{Implications for modeling coarse particle transport in the atmospheric surface layer}

Interpreting DNS results in terms of the laboratory or field scales must be done with care, as the Reynolds numbers in DNS numbers are much smaller than those found at these scales. However, by scaling up the results in this work, we can gain valuable qualitative insights into the drag on inertial settling dust particles. For example, using estimates of turbulent dissipation for an atmospheric surface layer of roughly $10^{-3}$ $\mathrm{m^2/s^3}$, we can define a rough Kolmogorov timescale as $10^{-1}$ s. Thus, for quartz dust particles ($\rho_p = 2650$ $\mathrm{kg/m^3}$) that range between 30-100 $\mathrm{\mu m}$, we should expect a values of $\mathrm{St}_\eta$ to range between 0.1--10. We can see that the ranges in our DNS are in the correct neighborhood to model these same coarse dust particles.
Moreover, we can use the values of $g^+$ from table \ref{tab:cases} (recall $g^+ = \mathrm{Sv^+/St^+}$) to estimate an equivalent friction velocity, $u_\tau^*$ (note that since we are re-scaling, $u_\tau*$ is necessarily different than the value of $u_\tau$ used in this work), which effectively gives us a qualitative estimate of the intensity of the turbulence in an atmospheric surface layer. The effective friction velocity is given by 
\begin{linenomath*}
    \begin{equation}
        u_\tau^* = \left(\frac{g\nu}{g^+}\right)^{1/3}.
    \end{equation}
\end{linenomath*}
Assuming $g = 9.81\mathrm{\,m/s^2}$ and $\nu = 1.57 \times 10^{-5}\mathrm{\,m^2/s}$, and some relationship between 10 metre wind velocity and the friction velocity (see \citet{kantha_small_2000} for example), this gives us a proxy for wind speed at the field scale. For the values of $g^+$ in this manuscript (see table \ref{tab:cases}), the effective friction velocities vary between 0.09 $\mathrm{m/s}$ and 0.84 $\mathrm{m/s}$, which covers a wide range of friction velocities on Earth \citep{vickers_formulation_2015}. 
Since $g^+$ is proportional to $\mathrm{Sv}^+$, we can see the effective wind speed increases as $\mathrm{Sv}^+$ decreases. 

Therefore, the insight we can gain is that the slip velocity in high wind conditions (small $\mathrm{Sv}^+)$ should be primarily governed by the fluctuations associated with the turbulence, as opposed to the mean induced by gravitational settling and the presence of the solid boundary. Conversely, at lower wind speeds, the drag induced by turbulent fluctuations is much smaller relative to the mean slip. Thus, the magnitude of the slip velocity should instead be controlled by the average, which itself is controlled by the Stokes settling velocity and the turbophoretic term. Likewise, we expect the higher moment terms governing the slip variance (i.e. $R_t$ and $R_g$) to be more important to the dynamics further away from the surface in this limit, relatively speaking. 

This is significant when applying models like BC2021 to particle transport in field scale systems. For example, as we have described previously, BC2021 can be used (in conjunction with a model for the unconditional fluid velocity variance) to predict the slip velocity variance for inertial settling particles in homogeneous turbulence. Under low $\mathrm{Sv}^+$ conditions, our results show that the magnitude of the slip velocity is primarily governed by its fluctuating component, which is in turn associated with the interactions with the turbulence. Moreover, as we have discussed, a locally homogeneous approximation may be used when $\mathrm{Sv}^+$ is small enough (see \eqref{inequality}), our work suggests that BC2021 can also be applied to inhomogeneous turbulence, like that of the atmospheric surface layer, provided we are not concerned with dynamics too close to the ground and the wind conditions are strong enough. Another interesting related application is towards modeling the particle Reynolds number, which is known to affect the associated drag on the particles (\citet{balachandar_scaling_2009}, Berk \& Coletti 2024 (under review)). For example, it is known that loitering effects are typically associated with large particle Reynolds number \citep{rosa_settling_2016}, and these loitering effects work to reduce the average particle settling velocity \cite{good_settling_2014}. Accurate modeling of loitering effects could explain discrepancies between numerical simulations and laboratory experiments with respect to the measurement of settling velocities \citep{ferran_experimental_2023}. Moreover, our results may gain some insights into further than expected horizontal transport of giant dust particles off of the West African Coast \citep{van_der_does_mysterious_2018}, which could be linked to loitering effects.

\section{Acknowledgements}
The authors would like to acknowledge Grant No. W911NF2220222 from the U.S. Army
Research Office. The authors would also like to thank the Center for Research Computing at the University
of Notre Dame.

\input{appendixA}
\input{appendixB}
\input{AppendixC}

\bibliographystyle{jfm}
\bibliography{zLibrary}

\end{document}

%% file: appendixA.tex
\section*{Appendix A -- Mathematical details of slip velocity model hierarchy}
By including gravitational settling in the particle equation of motion, there is now a mean settling velocity. Due to preferential sweeping, the average particle settling velocity can be increased (or decreased in some cases) beyond the laminar settling velocity, $v_g$, leading to there is a non-zero average slip velocity. Since we know that the average settling velocity of the particles will be non-zero due to the presence of gravity, the average of the squared mean is not equivalent to the average, i.e. $\langle F^2\rangle \neq \langle {F'}^2\rangle$, ($F$ is some arbitrary quantity, and a prime indicates a fluctuation about the mean of $F$) meaning we must be careful to  discern between the variance and squared means:
\begin{linenomath*}
    \begin{eqnarray}
      &\langle w_p^2\rangle_z = \langle w_p\rangle_z^2 + \langle {w_p'}^2\rangle_z, \\ 
    &\langle u_s^2\rangle_z = \langle u_s\rangle_z^2 + \langle {u_s'}^2\rangle_z,\quad \langle u_s\rangle_z = \langle u_f\rangle_z - \langle w_p\rangle_z, \\ 
      &\langle w_p^3\rangle_z = \langle w_p\rangle_z^3 + \langle w_p\rangle_z\langle {w_p'}^2\rangle_z + \langle {w_p'}^3\rangle_z. 
      \end{eqnarray}
\end{linenomath*}

Eq. \eqref{cons variance}, derived in \cite{johnson_turbophoresis_2020}, assumed that particles did not settle under the action of gravity, meaning that $\langle w_p\rangle_z = 0$. However, by substituting in the above Reynolds decompositions, it can be readily extended to settling particles.
Doing this, we arrive at
\begin{linenomath*}
    \begin{equation}
        \langle {u_s'}^2\rangle_z  = \langle {u_f'}^2\rangle_z - \langle {w_p'}^2\rangle_z-\frac{\tau_p}{\varrho}\frac{d}{dz}\varrho \langle w_p^3\rangle_z + \langle u_f\rangle_z^2 -\langle u_s\rangle_z^2 - \langle w_p\rangle_z^2 - 2\langle w_p\rangle_zv_g, \label{slip var 1}
    \end{equation}
\end{linenomath*}
where we have not expanded $\langle w_p^3\rangle_z$ in terms of its variance and mean components yet. We can see from this equation that the slip velocity variance is due to the variance of seen velocities, the variance of the particle velocity, and several other terms. These terms are difficult to interpret in their current form, so we simplify them next. 

The mean slip velocity squared, $\langle u_s\rangle_z^2$ can be expanded as
\begin{linenomath*}
    \begin{equation}
        \langle u_s\rangle_z^2 =\langle u_f\rangle_z^2 -2\langle u_f\rangle_z\langle w_p\rangle_z +\langle w_p\rangle_z^2.
    \end{equation}
\end{linenomath*}
Upon substitution of the above into \eqref{slip var 1} and simplifying, we can express \eqref{slip var 1} as
\begin{linenomath*}
    \begin{equation}
                \langle {u_s'}^2\rangle_z  = \langle {u_f'}^2\rangle_z -\langle {w_p'}^2\rangle_z -\frac{\tau_p}{\varrho}\frac{d}{dz}\varrho \langle w_p^3\rangle_z + 2\langle w_p\rangle_z\left( \langle u_s\rangle_z -v_g\right)
    \end{equation}
\end{linenomath*}
We can now expand the third term on the right hand side of the above in order to express the slip velocity variance in terms of only means and variances of other quantities.
\begin{linenomath*}
    \begin{multline}
                \langle {u_s'}^2\rangle_z  = \overbrace{\underbrace{\langle {u_f'}^2\rangle_z -\langle {w_p'}^2\rangle_z}_{(1)} -\frac{\tau_p}{\varrho}\frac{d}{dz}\varrho \langle {w_p'}^3\rangle_z}^{(2)} \\ \underbrace{-\frac{\tau_p}{\varrho}\frac{d}{dz}\left(\varrho\langle w_p\rangle^3_z + 3\varrho\langle w_p\rangle_z\langle {w_p'}^2\rangle_z\right)
                + 2\langle w_p\rangle_z\left( \langle u_s\rangle_z -v_g\right)}_{(3)}
    \end{multline}
\end{linenomath*}


%% file: appendixB.tex
\section*{Appendix B -- Response function model of acceleration variance}
In this section, we sketch the derivation of the semi-analytical model proposed by \cite{berk_dynamics_2021} for the slip velocity variance. 
The following is based on the concept of a response function, described in \cite{csanady_turbulent_1963}, which is meant to quantify the fact that inertial particles require a finite amount of time to respond to turbulent fluctuations. By Fourier transforming the vertical component of the fluctuating particle velocity and the sampled velocity through the use of standard manipulations, we can write the slip velocity variance and the particle velocity variance as
\begin{linenomath*}
    \begin{equation}
        \langle {u_s^\prime}^2\rangle = \int_0^{\infty}(\omega\tau_p)^2 E_pd\omega, \quad \langle {w_p^\prime}^2\rangle = \int_0^{\infty}E_pd\omega.
    \end{equation}
\end{linenomath*}
In this equation $E_p$ is the kinetic energy spectrum of the particles. As discussed in \cite{csanady_turbulent_1963}, $E_p$ is related to the kinetic energy spectrum of the fluid motion sampled along the particle's trajectory, denoted by $E_{f,p}$, through a response function
\begin{equation}
    H(\omega) = \frac{1}{1 + \left(\omega\tau_p\right)^2},
\end{equation} meaning that
\begin{linenomath*}
    \begin{equation}
        \langle {u_s^\prime}^2\rangle = \int_0^{\infty}(\omega\tau_p)^2H(\omega) E_{f,p}d\omega,\quad \langle {w_p^\prime}^2\rangle = \int_0^{\infty}H(\omega) E_{f,p}d\omega. 
    \end{equation}
\end{linenomath*}
Using the stochastic model for the particle velocity auto-correlation outlined in \cite{sawford_reynolds_1991}, \citet{berk_dynamics_2021} used the fact that the auto-correlation and the spectra are fourier transform pairs. The particle velocity auto-correlation described in \cite{sawford_reynolds_1991} is 
\begin{linenomath*}
    \begin{equation}
        R_{f,p}(t) = \frac{\langle {u_f^\prime}^2\rangle}{\tau_{l,p} - \tau_2}\left(\tau_{l,p}e^{-t/\tau_{l,p}}+\tau_{2}e^{-t/\tau_{2}}\right),
    \end{equation}
\end{linenomath*}
where $\tau_2$ is proportional to the fluid acceleration variance and appears due to the finite Reynolds number. For this work, we use 
\begin{linenomath*}
    \begin{equation}
        \tau_2 = \frac{C_0}{2a_0}\tau_\eta,
    \end{equation}
\end{linenomath*}
where $C_0$ and $a_0$ are universal constants modeled by
\begin{linenomath*}
    \begin{equation}
        C_0 = C_\infty(1 - (0.1\mathrm{Re_\lambda}^{-1/2})), \quad a_0 = \left(\frac{5}{1 + 100\mathrm{Re}_\lambda^{-1}}\right),
    \end{equation}
\end{linenomath*}
defined in \citet{lien_kolmogorov_2002} and \citet{sawford_conditional_2003} respectively. Here, $\mathrm{Re}_\lambda = \sqrt{15}\langle w^2\rangle/v_\eta^2$ is the Taylor Reynolds number evaluated at a height $z$. Thus, $C_0$ and $a_0$ are functions of the vertical coordinate. Note that the results in this work are not meaningfully dependent on the exact choice of model for $C_0$ and $a_0$.

$\tau_{l,p}$ is the lagrangian correlation timescale of the turbulence along the particle trajectory. $\tau_{l,p}$ is a function of three parameters: ratio of the laminar settling velocity to the integral velocity scale, the lagrangian correlation timescale of the turbulence, and the Eulerian correlation timescale of the turbulence which are defined as
\begin{linenomath*}
    \begin{equation}
        \mathrm{Sv}_\ell =\frac{v_g}{w^\prime},\quad \tau_E = \frac{\langle w^2\rangle}{\epsilon}, \quad \tau = -\frac{\kappa z}{u_\tau}\frac{\langle uw\rangle}{\langle w^2\rangle},
    \end{equation}
\end{linenomath*}
respectively. Note that the definition of $\tau$ can be found in \citet{oesterle_lagrangian_2004}.
$\tau_{l,p}$ is meant to encapsulate the fact that as an inertial particle settles through a local neighborhood of correlated motion, the turbulence it experiences de-correlates faster along its trajectory than it would if it was not settling. \cite{berk_dynamics_2021} derived a semi-empirical model for the correlation along the particle trajectory using the idea of the crossing trajectories mechanism introduced by \cite{csanady_turbulent_1963} as:
\begin{linenomath*}
    \begin{equation}
        \tau_{l,p} = \tau\frac{1}{\left(1 + \left(\frac{\tau}{\tau_E}\right)^2\mathrm{Sv}_\ell^2\right)^{1/2}}
    \end{equation}
\end{linenomath*}

By Fourier Transforming $R_{f,p}$ and substituting into the integral relations for the slip variance and velocity variance, we arrive at the following for the 
velocity variance
\begin{linenomath*}
    \begin{equation}
        \langle {w_p^\prime}^2\rangle = \langle {u_f^\prime}^2\rangle\left(1 - \frac{\mathrm{St}^2_\eta}{\left(\mathrm{St}_\eta + \frac{\tau_{l,p}}{\tau_\eta}\right)\left(\mathrm{St}_\eta + \frac{\tau_{2}}{\tau_\eta}\right)}\right).
        \label{vel var model}
    \end{equation}
\end{linenomath*}
and the slip variance
\begin{linenomath*}
    \begin{equation}
        \langle {u_s^\prime}^2\rangle = \langle {u_f^\prime}^2\rangle\frac{\mathrm{St}^2_\eta}{\left(\mathrm{St}_\eta + \frac{\tau_{l,p}}{\tau_\eta}\right)\left(\mathrm{St}_\eta + \frac{\tau_{2}}{\tau_\eta}\right)},
        \label{slip var model}
    \end{equation}
\end{linenomath*}
If follows from the above that the particle acceleration variance is
\begin{linenomath*}
    \begin{equation}
        \langle {a_p^\prime}^2\rangle = \langle {u_f^\prime}^2\rangle\frac{1}{\left(\mathrm{St}_\eta + \frac{\tau_{l,p}}{\tau_\eta}\right)\left(\mathrm{St}_\eta + \frac{\tau_{2}}{\tau_\eta}\right)},
        \label{accel var model}
    \end{equation}
\end{linenomath*}

Since the term in the brackets in \eqref{vel var model} is simply the model for $1 - \langle {u_s^\prime}^2\rangle/\langle {u_f^\prime}^2\rangle$, the implied relationship between the slip variance and the particle velocity variance is 
\begin{linenomath*}
\begin{equation}
\langle {u_s^\prime}^2\rangle = \langle {u_f^\prime}^2\rangle - \langle {w_p^\prime}^2\rangle, 
\end{equation}   
\end{linenomath*}
which is almost identical to \eqref{cons slip var}, except for the fact that $R_t$ and $R_g$ are not accounted for in this model.


%% file: AppendixC.tex
\section*{Appendix C -- Comparison of computed and modeled slip variance}

\begin{figure}
    \centering
    \includegraphics[width=\textwidth]{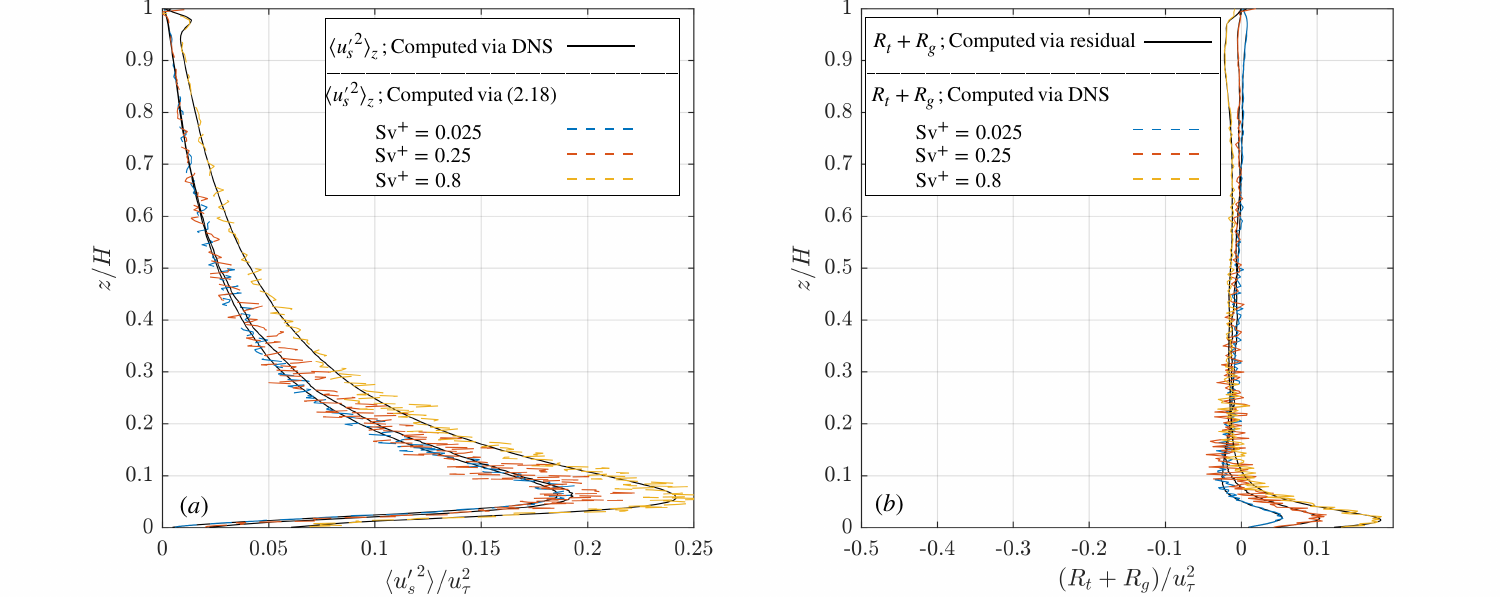}
    \caption{Panel (a) shows the slip velocity variance computed by the DNS for three values of $\mathrm{Sv}^+$ at fixed $\mathrm{St}^+=10$ (black curves), and the slip variance computed by \eqref{cons slip var} (dashed colored curves). Panel(b) shows $R_t + R_g$ computed via a residual of $\langle {u_s^\prime}\rangle_z - \left(\langle {u_f^\prime}\rangle_z - \langle{w_p^\prime}\rangle_z\right)$ (black curves), and $R_t+R_g$ computed directly (dashed colored curves) for the same cases.}
    \label{fig:appC}
\end{figure}

Figure \ref{fig:appC}(a) shows a comparison between computed values of $\langle {u_s^\prime}^2\rangle_z$ and that computed using the right hand side of \eqref{cons slip var}. The slip variance computed directly from the DNS for three different values of $\mathrm{Sv}^+$ at $\mathrm{St}^+=10$ are shown by black curves, while the right hand side of \eqref{cons slip var} are shown by colored dashed lines. We can see that the right hand side contains significant noise, but otherwise models the computed slip variance well. This noise is a result of the routines used to estimate the derivatives in $R_t$ and $R_g$, and not in the computation of $\langle {u_f^\prime}^2\rangle_z$ and $\langle {w_p^\prime}^2\rangle_z$ (as evidenced in figure \ref{fig:var profs}).
Figure \ref{fig:appC}(b) shows a comparison between the computed values of $R_t+R_g$ (dashed curves) and that computed by a residual of \eqref{cons slip var}(black solid curves). We can see that by plotting $R_t$ and $R_g$ as a residual, the noise is significantly reduced.

